\documentclass[lettersize,journal]{IEEEtran}

\usepackage{float}
\usepackage{color}
\usepackage{graphicx}
\usepackage{mathtools}
\usepackage{amsmath,amssymb,amsthm}

\usepackage[caption=false,subrefformat=parens,labelformat=parens]{subfig}
\usepackage{balance}
\usepackage[noadjust]{cite}
\usepackage{epstopdf}
\usepackage{psfrag,pst-pdf}
\usepackage[nolist]{acronym}

\usepackage{algcompatible}
\usepackage{algorithm}
\usepackage{dsfont}
\usepackage{textgreek}
\usepackage{threeparttable}

\usepackage{algpseudocode}

\makeatletter
\newcounter{phase}[algorithm]
\newlength{\phaserulewidth}

\makeatother

\usepackage{hyperref}

\acrodef{2D}{two-dimensional}
\acrodef{3D}{three-dimensional}
\acrodef{4G}{4th generation}
\acrodef{5G}{5th generation}
\acrodef{6G}{sixth generation}
\acrodef{ADT}{angle diversity transmitter}
\acrodef{ADR}{angle diversity receiver}
\acrodef{ANSI}{American national standards institute }
\acrodef{AP}{access point}
\acrodef{AR}{augmented reality}
\acrodef{AWGN}{additive white Gaussian noise}
\acrodef{AWGR}{arrayed waveguide grating router}
\acrodef{BER}{bit error ratio}
\acrodef{CDF}{cumulative distribution function}
\acrodef{CSI}{channel state information}
\acrodef{CPC}{compound parabolic concentrator}
\acrodef{CoMB}{coordinated multi-beam}
\acrodef{CoMB{-}JT}{coordinated multi-beam joint transmission}
\acrodef{CoMP}{coordinated multi-point}
\acrodef{DC}{direct current}
\acrodef{DCO{-}OFDM}{direct current-biased optical orthogonal frequency division multiplexing}
\acrodef{DD}{direct detection}
\acrodef{DFB}{distributed feedback laser}
\acrodef{DMT}{discrete multitone}
\acrodef{EE}{energy efficiency}
\acrodef{EGC}{equal gain combining}
\acrodef{FEC}{forward error correction}
\acrodef{FFT}{fast Fourier transform}
\acrodef{FF}{fill factor}
\acrodef{FOV}{field of view}
\acrodef{FSO}{free-space optical}
\acrodef{IBI}{inter-beam interference}
\acrodef{IEC}{International electrotechnical commission}
\acrodef{IFFT}{inverse fast Fourier transform}
\acrodef{IM}{intensity modulation}
\acrodef{IM{/}DD}{intensity modulation and direct detection}
\acrodef{IM{-}DD}{intensity modulation and direct detection}
\acrodef{IR}{infrared}
\acrodef{ISI}{inter-spot interference}
\acrodef{ICI}{inter-cluster interference}
\acrodef{JT}{joint transmission}
\acrodef{LD}{laser diode}
\acrodef{LED}{light-emitting diode}
\acrodef{LiFi}{light fidelity}
\acrodef{MIMO}{multiple input multiple output}
\acrodef{MMF}{multi-mode fiber}
\acrodef{MPE}{maximum permissible exposure}
\acrodef{MPTP}{maximum permissible transmit power}
\acrodef{MHP}{most hazardous position}
\acrodef{MRC}{maximum ratio combining}
\acrodef{MUI}{multi-user interference}
\acrodef{NIR}{near infrared}
\acrodef{NOMA}{non-orthogonal multiple access}
\acrodef{OOK}{on-off keying}
\acrodef{PAM}{pulse amplitude modulation}
\acrodef{PSD}{power spectral density}
\acrodef{PD}{photodiode}
\acrodef{OFDM}{orthogonal frequency division multiplexing}
\acrodef{OFDMA}{orthogonal frequency division multiple access}
\acrodef{OCC}{optical camera communication}
\acrodef{OW}{optical wireless}
\acrodef{OWC}{optical wireless communication}
\acrodef{QAM}{quadrature amplitude modulation}
\acrodef{RF}{radio frequency}
\acrodef{RIN}{relative intensity noise}
\acrodef{RSS}{received signal strength}
\acrodef{SDMA}{space division multiple access}
\acrodef{SIC}{successive interference cancellation}
\acrodef{SiPM}{silicon photomultiplier}
\acrodef{SMF}{single-mode fiber}
\acrodef{SINR}{signal-to-interference-plus-noise ratio}
\acrodef{SNR}{signal-to-noise ratio}
\acrodef{Tb{/}s}{Terabit/s}
\acrodef{TIA}{transimpedance amplifier }
\acrodef{TEM}{transverse electromagnetic}
\acrodef{UHD}{ultra-high-definition}
\acrodef{VCSEL}{vertical cavity surface emitting laser}
\acrodef{VLC}{visible light communication}
\acrodef{VR}{virtual reality}
\acrodef{WiFi}{wireless fidelity}

\acrodef{ADC}{analog-to-digital converter}
\acrodef{DAC}{digital-to-analog converter}
\acrodef{CPU}{central processing unit}
\acrodef{BS}{base station}
\acrodef{ETP}{effective transmit power}
\acrodef{LIV}{light-current-voltage}
\acrodef{PA}{power amplifier}
\acrodef{PC}{power consumption}
\acrodef{UE}{user equipment}

\acrodef{GoB}{grid-of-beam}
\acrodef{SBC}{static beam clustering}
\acrodef{DBC}{dynamic beam clustering}
\acrodef{CMOS}{complementary metal-oxide semiconductor}
\acrodef{DSP}{digital signal processing}
\acrodef{GOPS}{giga operations per second}
\acrodef{GSPS}{giga samples per second}
\acrodef{FinFET}{fin field-effect transistor}
\acrodef{FOM}{figure of merit}
\acrodef{SNDR}{signal-to-noise-and-distortion ratio}
\acrodef{ENOB}{effective number of bits}
\acrodef{AC}{alternating current}
\acrodef{SRAM}{static random access memory}
\acrodef{FPGA}{field-programmable gate array}
\acrodef{CPE}{computational power efficiency}
\acrodef{QoS}{quality of service}
\acrodef{BFS}{breadth-first search}
\acrodef{BDMA}{beam division multiple access}
\acrodef{AI}{artificial intelligence}
\acrodef{IoT}{Internet of Things}
\acrodef{TV}{television}

\acrodef{SDI}{serial digital interface}
\acrodef{NRZ}{non-return-to-zero}
\acrodef{OOK}{on-off keying}
\acrodef{NRZ{-}OOK}{non-return-to-zero on-off keying}
\acrodef{BPSK}{binary phase-shift keying}

\acrodef{MM}{multi-mode}
\acrodef{SM}{single-mode}
\acrodef{CP}{cyclic prefix}
\acrodef{CLT}{central limit theorem}
\acrodef{FF}{far field}
\acrodef{FWHM}{full-width at half-maximum}
\acrodef{AWG}{arbitrary waveform generator}
\acrodef{SPS}{samples per symbol}
\acrodef{BT}{bias tee}
\acrodef{IL}{insertion loss}
\acrodef{ACL}{aspheric condenser lens}
\acrodef{NA}{numerical aperture}
\acrodef{CA}{clear aperture}
\acrodef{ARC}{anti-reflective coating}
\acrodef{FC{/}PC}{Ferrule Connector/Physical Contact}
\acrodef{VGA}{variable gain amplifier}
\acrodef{BLER}{block error rate}
\acrodef{LTE}{long term evolution}
\acrodef{HH}{Hughes-Hartogs}
\acrodef{MCS}{modulation and coding scheme}
\acrodef{HARQ}{hybrid automatic repeat request}
\acrodef{SMD}{surface‑mount device}
\acrodef{GaN}{gallium nitride}
\acrodef{WDM}{wavelength division multiplexing}
\acrodef{PIN}{positive-intrinsic-negative}
\acrodef{InGaAs}{indium gallium arsenide}
\acrodef{GaAs}{gallium arsenide}
\acrodef{VI}{Vertically Integrated}
\acrodef{VIS}{Vertically Integrated Systems}
\acrodef{CSI}{channel state information}
\acrodef{MA}{multi-aperture}
\acrodef{RMS}{root mean square}
\acrodef{OSA}{optical spectrum analyzer}
\acrodef{EMB}{effective modal bandwidth}
\acrodef{CD}{chromatic dispersion}
\acrodef{MD}{modal dispersion}
\acrodef{VL}{visible light}
\acrodef{LiDAR}{light detection and ranging}
\acrodef{EML}{electro-absorption modulated laser}
\acrodef{EDFA}{erbium-doped fiber amplifier}
\acrodef{EEL}{edge emitting laser}
\acrodef{Si}{silicon}
\acrodef{EFL}{effective focal length}
\acrodef{AFC}{achromatic fiber collimator}
\acrodef{DOF}{degrees of freedom}
\acrodef{PWL}{piecewise linear}

\begin{document}

\title{Achieving 70 Gb/s Over A VCSEL-Based Optical Wireless Link Using A Multi-Mode Fiber-Coupled Receiver}

\author{\IEEEauthorblockN{Hossein~Kazemi,~\IEEEmembership{Member,~IEEE}, Isaac~N.~O.~Osahon,~\IEEEmembership{Member,~IEEE}, Nikolay~Ledentsov~Jr., Ilya~Titkov, Nikolay~Ledentsov,~\IEEEmembership{Senior Member,~IEEE}, and~Harald~Haas,~\IEEEmembership{Fellow,~IEEE}\vspace{-10pt}}
\thanks{
	\indent Hossein Kazemi \textit{(corresponding author)}, Isaac N. O. Osahon, and Harald Haas are with the LiFi Research and Development Center (LRDC), Electrical Engineering Division, Department of Engineering, University of Cambridge, Cambridge CB3 0FA, United Kingdom. \\
    \indent Nikolay Ledentsov Jr., Ilya Titkov, and Nikolay Ledentsov are with VI Systems GmbH (VIS), Berlin 10623, Germany.}}
\maketitle

\begin{abstract}
In this paper, we demonstrate a laser-based optical wireless communication (OWC) system employing a $940$~nm single-mode (SM) vertical cavity surface emitting laser (VCSEL) and a multi-mode (MM) fiber-coupled receiver, achieving a record data rate beyond $70$~Gb/s, while the optical transmit power is below $5$~mW. The use of a high speed fiber-optic photoreceiver avoids limiting the communication bandwidth by the receiver, enabling ultra-high capacity and energy-efficient light fidelity (LiFi) links to unlock new applications. This work experimentally validates the feasibility of ultra-high speed indoor OWC systems using a single low-power and low-cost VCSEL for next-generation LiFi connectivity.

\end{abstract}

\begin{IEEEkeywords}
Next-generation light fidelity (LiFi), laser-based optical wireless communication (OWC), ultra-high speed, vertical cavity surface emitting laser (VCSEL), fiber-optic photoreceiver.
\vspace{-10pt}
\end{IEEEkeywords}

\section{Introduction} \label{Sec:Introduction}
The next generation of \ac{LiFi} technology is set to leverage laser-based \ac{OWC} to overcome the bandwidth limitations of conventional \ac{LED}-based \ac{OWC} systems and meet the stringent requirements of \ac{6G} networks \cite{Chow2024OWC6G}. Laser-based \ac{OWC} offers a cost-effective solution by enabling ultra-high speed and energy-efficient wireless links to support real-time bandwidth-intensive applications in \ac{6G} targeting multi-Gb/s data rates with ultra-low latency, such as \ac{AR}, \ac{VR}, 4K and 8K \ac{UHD} video streaming, ultra-high data density services, and inter-rack connections in data centers \cite{Chow2024OWC6G,Schrenk2025Wideband,Ali2019PTL}.

Achieving data rates of tens of Gb/s using laser-based \ac{OWC} for indoor applications has been actively investigated \cite{CLee2022_26Gbps,CChen2024100Gbps,Osahon2024VCSEL38Gbps}. In \cite{CLee2022_26Gbps}, Lee~\textit{et al.} experimentally validated a dual‑wavelength \ac{WDM} \ac{OWC} system using a novel white‑light transmitter by integrating a $450$~nm \ac{GaN} blue laser, an $850$~nm \ac{IR} laser, and a phosphor converter within a single \ac{SMD} package. The authors reported a combined data rate exceeding $26$~Gb/s over a $3$~m link at a target \ac{BER} of $5.6\times10^{-2}$ adopted from \ac{LTE} systems, using two \ac{Si} \ac{PIN} photoreceivers with a $3$~dB bandwidth of $1.4$~GHz to detect the optical signals of the blue and \ac{IR} \acp{LD}. In \cite{CChen2024100Gbps}, Chen~\textit{et al.} implemented a $10$‑channel \ac{LiFi} link using the same \ac{SMD} laser technology based on a transmitter setup that integrates ten high‑brightness, high‑bandwidth \acp{LD}, including three \ac{GaN} blue \acp{LD} operating at $405$~nm, $450$~nm, and $455$~nm, and seven \ac{IR} \acp{LD} with wavelegths of $850$~nm to $1064$~nm. To detect the optical signals, $1.4$~GHz \ac{Si} \ac{PIN} photoreceivers were used, except for the $1064$~nm \ac{LD}, for which a $2$~GHz \ac{InGaAs} photoreceiver was used due to its higher responsivity. The authors demonstrated aggregate data rates in excess of $100$~Gb/s over a few meter link distance by considering a \ac{BER} threshold of $5.6\times10^{-2}$.

Among various types of \acp{LD}, \acp{VCSEL} are well-suited for realizing laser-based \ac{OWC} due to their outstanding features. They are low-cost and low-power devices (e.g. with a few mW power consumption), and have an extensive modulation bandwidth, well-controlled output beam properties, enhanced reliability, and the ability to be fabricated as \ac{2D} arrays \cite{NLedentsov2022HighSpeed}. \Ac{GaAs}-based \acp{VCSEL} are mainly established in the \ac{NIR} wavelength region, including $850$~nm for data communications and $940$--$980$~nm for sensing applications \cite{ALiu2019VCSELs}. Mature $850$~nm \acp{VCSEL} provide a maximum output power around $10$~mW and a modulation bandwidth $>10$~GHz, suitable for short-reach \ac{MMF} links and low-complexity \ac{OWC} systems based on \ac{Si} \acp{PD} \cite{CHCheng2018VCSEL}. $940$~nm \acp{VCSEL} have similar power and bandwidth specifications while offering improved eye safety and reduced solar background, which is desired for \ac{3D} sensing technologies such as face and gesture recognition, and \ac{LiDAR} \cite{CHCheng2018VCSEL}. These unique features make \acp{VCSEL} a strong contender for short-range, ultra-high speed and energy-efficient \ac{OWC} systems for next-generation wireless connectivity \cite{Kazemi2025GoB}. Furthermore, \ac{SM} \acp{VCSEL} emit a Gaussian laser beam with a narrow linewidth and offer a lower \ac{FF} divergence angle compared to their \ac{MM} counterparts \cite{RMichalzik2012}. This leads to a smaller beam spot size and improved light coupling efficiency when the beam is focused on the small active area of a high speed \ac{PD}, which is indeed an advantage for high speed point-to-point optical wireless link design. In \cite{Osahon2024VCSEL38Gbps}, Osahon~\textit{et al.} experimentally demonstrated a line‑of‑sight optical wireless link based on an $850$~nm \ac{SM} \ac{VCSEL} with a $3$-dB bandwidth of $6.2$~GHz, achieving a gross data rate of $38$~Gb/s over $2.5$~m with an eye‑safe launch power of $–1.47$~dBm (i.e., $0.7$~mW), assuming a $7\%$ \ac{FEC} limit for a pre-\ac{FEC} \ac{BER} of $3.8\times10^{-3}$. This performance was achieved by using a $9$~GHz \ac{Si} \ac{PIN} \ac{PD} and spreading the data across a large number of \ac{DMT} subcarriers by means of adaptive bit and power loading.

In this work, we demonstrate a laser-based optical wireless link experiment using a $940$~nm \ac{SM} \ac{VCSEL}, reaching data rates of $>70$~Gb/s with an average optical power of $<5$~mW. The \ac{VCSEL} has an ultra-wide modulation bandwidth in excess of $18$~GHz. In order to realize the full potential of this \ac{VCSEL}, we employ a $25$~GHz fiber-coupled receiver to ensure that the receiver does not limit the end-to-end bandwidth of the system. In addition, we apply \ac{DCO{-}OFDM} with adaptive bit and power loading. To the best of our knowledge, this is the first experimental proof of a directly modulated \ac{SM} \ac{VCSEL} at $940$~nm in conjunction with a \ac{MM} fiber-optic photoreceiver for ultra-high speed laser-based \ac{OWC}, achieving a record data rate of $72$~Gb/s using a single low-power \ac{VCSEL}. This brings about a fourfold increase in the achievable data rate compared to our previous work \cite{HKazemi2023Experimental}.

The rest of the paper is organized as follows. In Section~\ref{Sec:ExperimentalSetup}, the experimental setup of the proposed \ac{VCSEL}-based \ac{OWC} system is described. In Section~\ref{Sec:Results}, the measurement results are discussed, in terms of the frequency response of the end-to-end system, the adaptive bit and power loading performance, the achievable data rate and the \ac{BER} performance. In Section~\ref{Sec:Conclusions}, concluding remarks and future works are presented.


\begin{figure}[!t]
	\centering
	\includegraphics[width=0.9\linewidth]{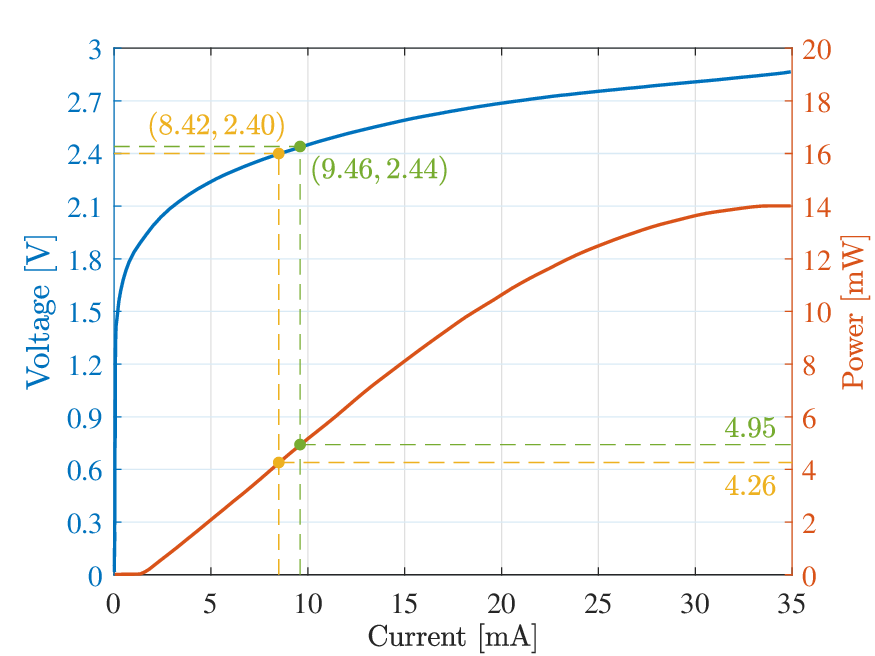}
	\caption{LIV characteristic curves of the MA VCSEL measured at $25^\circ$C. The DC operating points used in the experiment are shown by dashed green and yellow lines; see Table~\ref{Tab:DC_Bias_Configs}.}
    \label{Fig:VCSEL_LIV}
\end{figure}

\section{Experimental Setup} \label{Sec:ExperimentalSetup}
\subsection{VCSEL Characterization}
We use an \ac{SM} \ac{VCSEL} operating at a center wavelength of $940$~nm with a low spectral width. This \ac{VCSEL} was designed, manufactured and packaged by \ac{VIS} GmbH and was tested at the wafer level before being fully characterized for optical communications \cite{NLedentsov2022Advances}. It is a \ac{MA} \ac{VCSEL} featuring a compact mini-array structure, comprising four oxide-confined \acp{VCSEL}, each with an effective aperture diameter of about $7$~{\textmu}m, which are arranged in a $2\times2$ array configuration with a small pitch of $<15$~{\textmu}m \cite{NLedentsov2022Advances}. These four \acp{VCSEL} have a common pair of contacts such that they act as a single \ac{VCSEL}. Consequently, the output power of the \ac{MA} \ac{VCSEL} is effectively improved by a factor of four while maintaining both the \ac{SM} emission spectrum and the broadband modulation response of individual \acp{VCSEL} in the mini-array structure \cite{NLedentsov2022Advances}. The small-pitch \ac{VCSEL} mini-array has a compact footprint for efficient light coupling into a $50$~{\textmu}m core \ac{MMF} \cite{NLedentsov2022HighSpeed}. Such a miniature structure is fabricated using oxidation through holes \cite{NLedentsov2022HighSpeed}. A die shot of the \ac{VCSEL} mini-array chip is available in \cite{NLedentsov2022Advances}.

\begin{figure}[!t]
	\centering
	\includegraphics[width=0.9\linewidth]{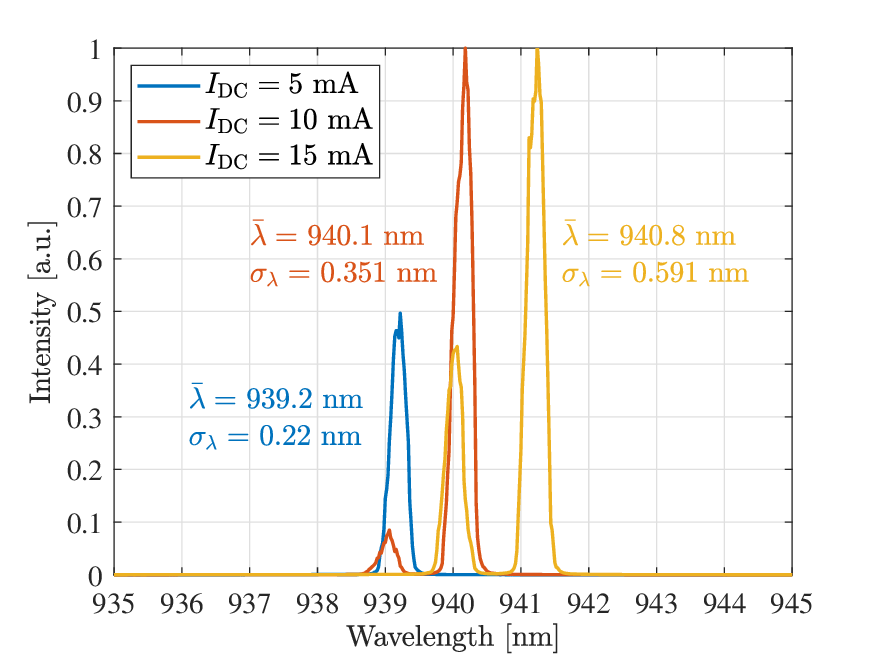}
	\caption{Optical spectrum of the MA VCSEL measured at different DC bias currents. The average wavelength $\bar{\lambda}$ and the RMS spectral width $\sigma_\lambda$ for each case are provided next to the corresponding optical spectrum.}
    \label{Fig:VCSEL_OpticalSpectrum}
\end{figure}

\begin{figure}[!t]
    \centering
    \subfloat[$I_\mathrm{DC}=5$ mA \label{Fig:FarFieldIntensity_5mA}]{\includegraphics[width=0.31\linewidth, keepaspectratio=true]{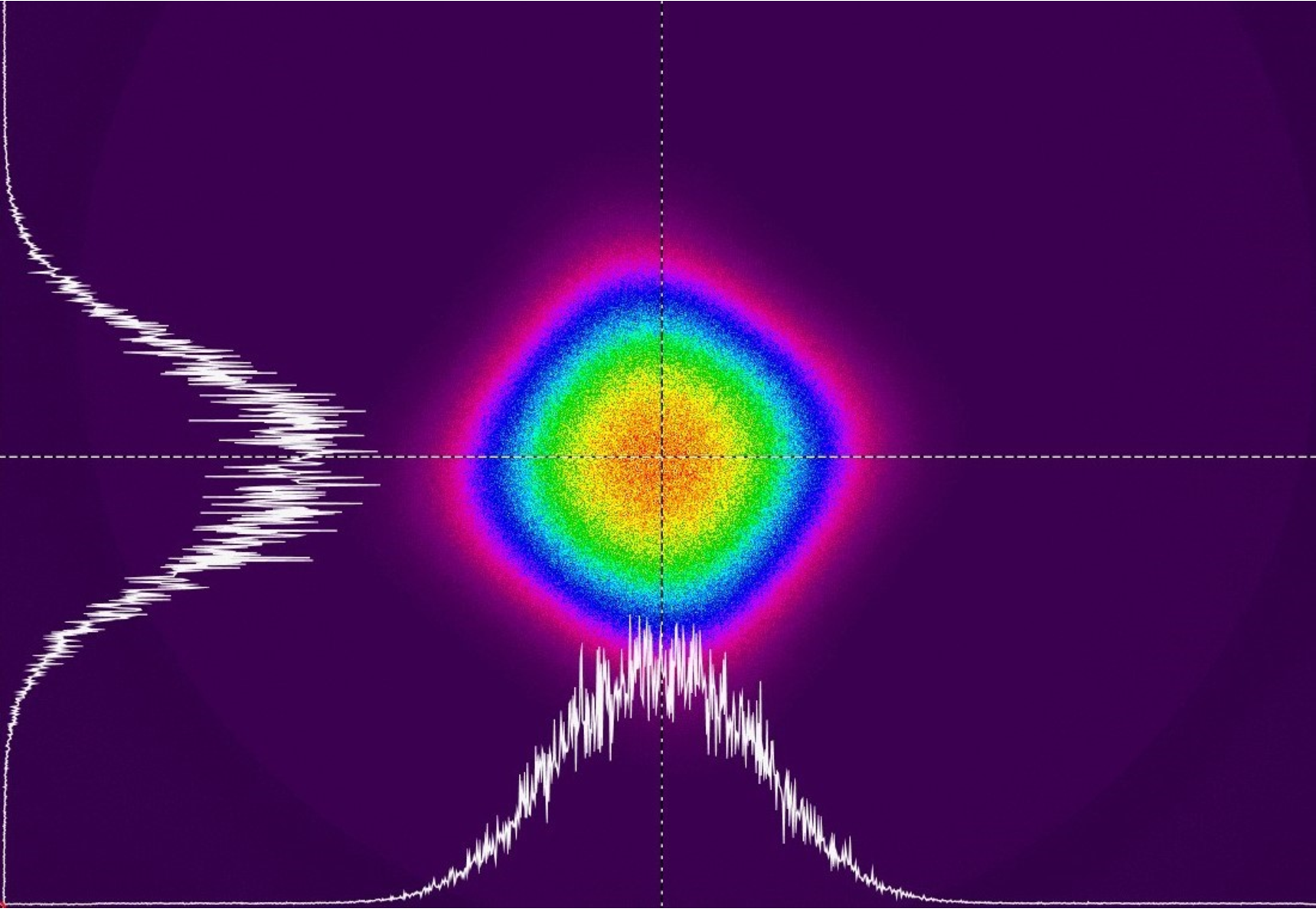}} \hspace{5pt}
    \subfloat[$I_\mathrm{DC}=10$ mA \label{Fig:FarFieldIntensity_10mA}]{\includegraphics[width=0.31\linewidth, keepaspectratio=true]{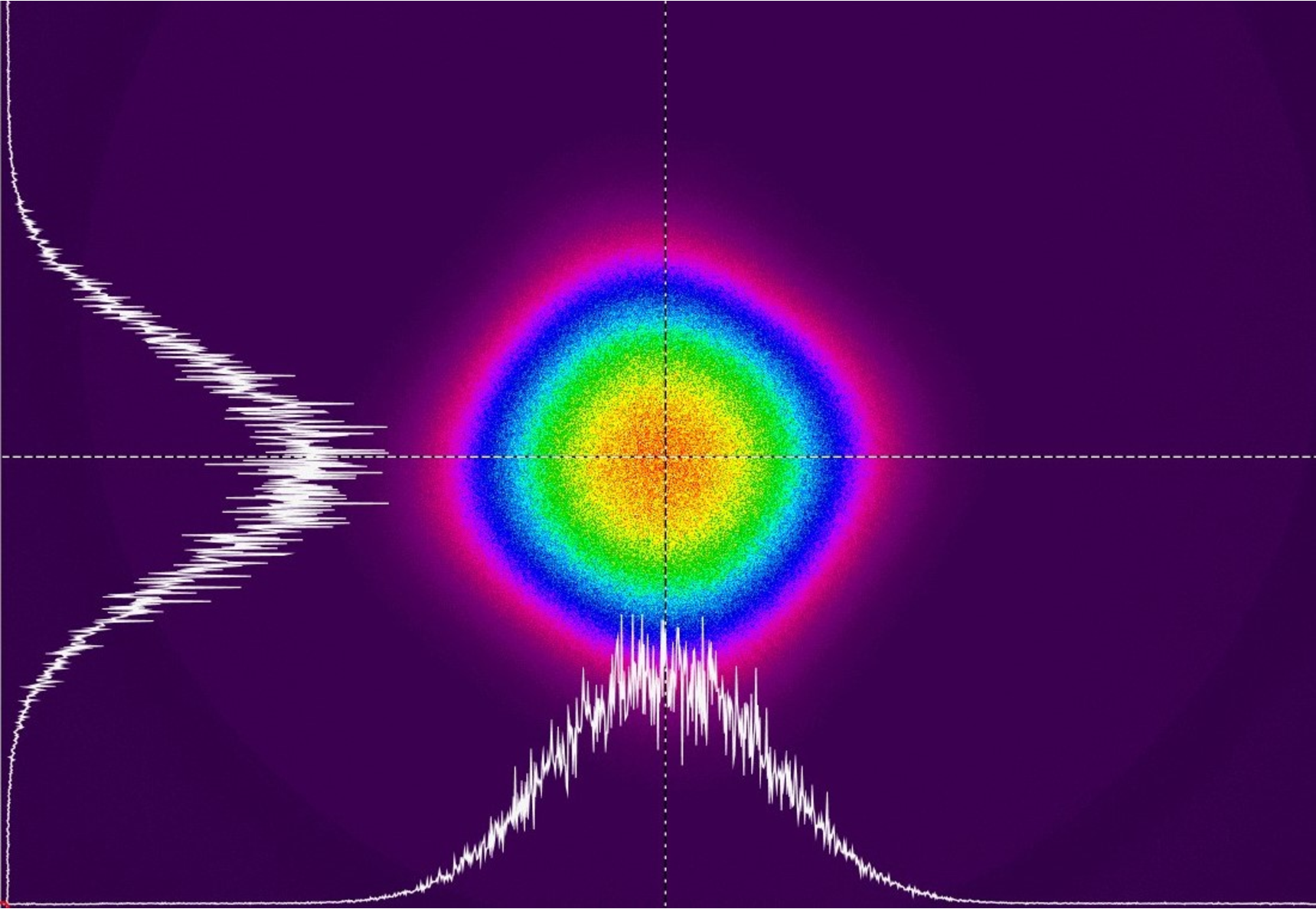}} \hspace{5pt}
    \subfloat[$I_\mathrm{DC}=15$ mA \vspace{3pt} \label{Fig:6_NoClustering_15mA}]{\includegraphics[width=0.31\linewidth, keepaspectratio=true]{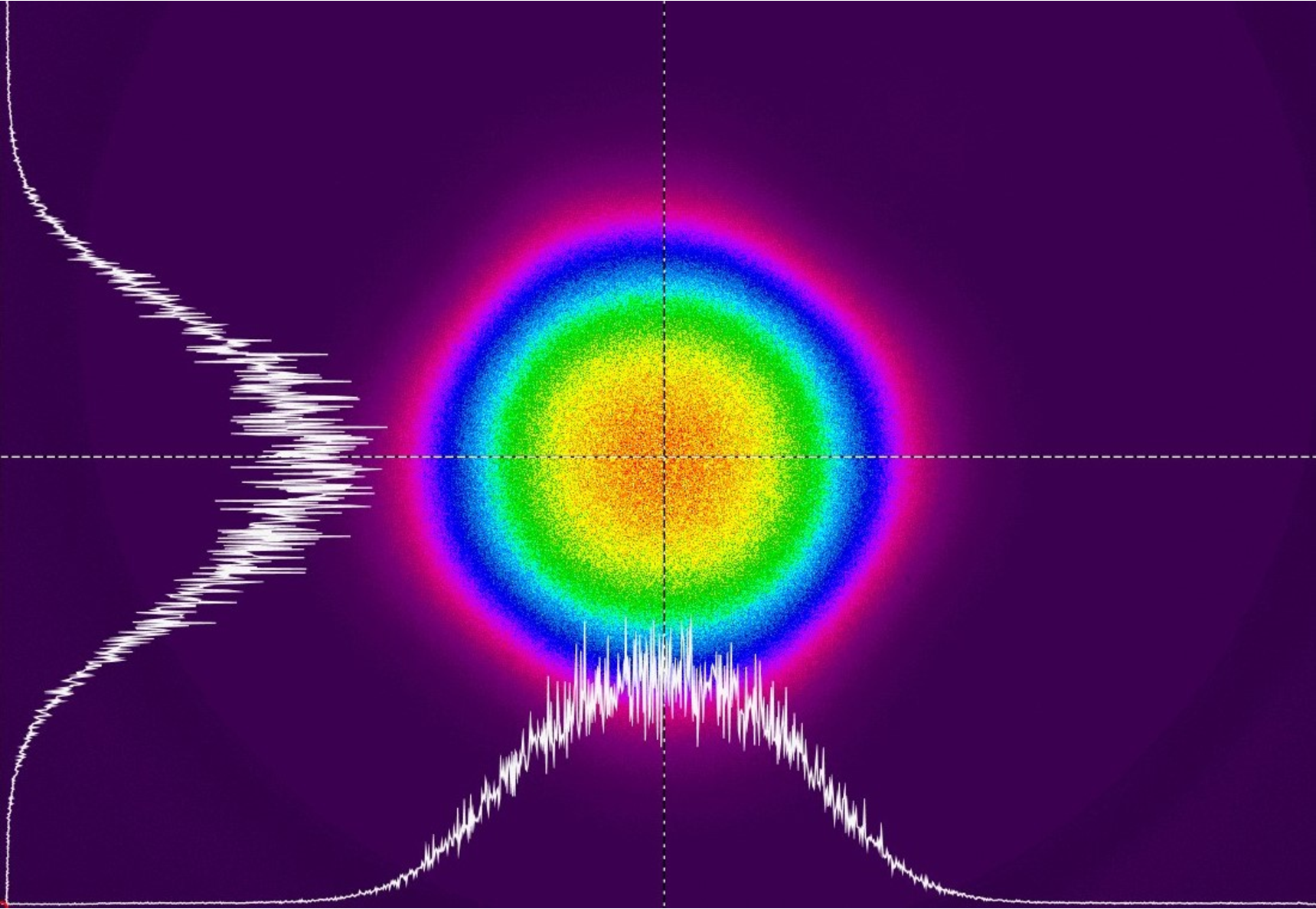}}
    \caption{FF intensity profile of the MA VCSEL measured at different DC bias currents. The FWHM divergence angle varies between $15^\circ$ and $18^\circ$.}
    \label{Fig:VCSEL_FarFieldIntensity}
\end{figure}

\newcounter{mycounter1}
\setcounter{mycounter1}{\value{figure}}
\setcounter{figure}{4}
\begin{figure*}[!t]
    \centering
    \includegraphics[width=\linewidth]{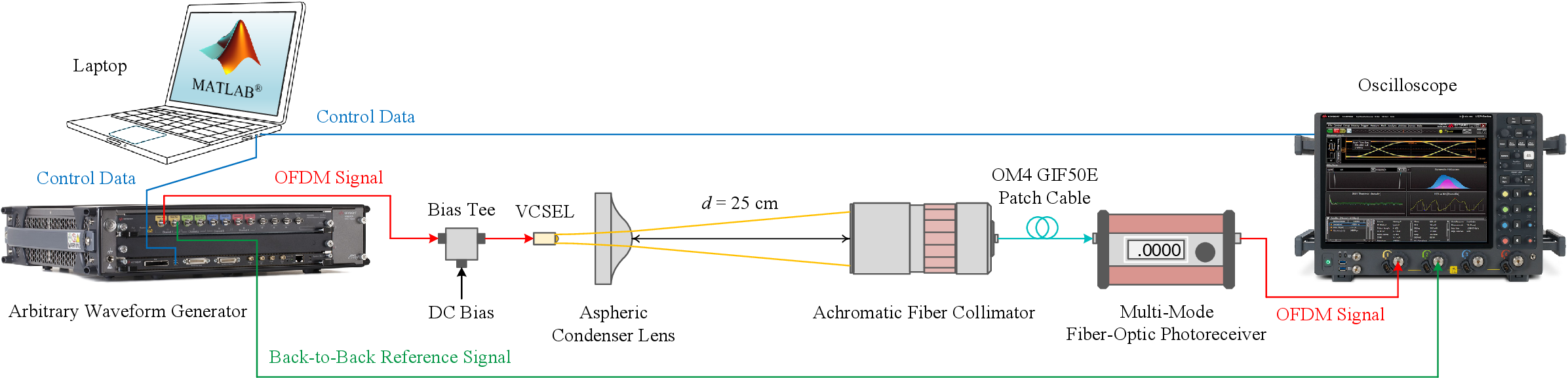}
    \caption{Ultra-high speed OWC system design based on a SM MA VCSEL and a MM fiber-optic photoreceiver using DCO-OFDM modulation.}
    \label{Fig:BlockDiagram}
\end{figure*}
\setcounter{figure}{\value{mycounter1}}

Fig.~\ref{Fig:VCSEL_LIV} shows the \ac{LIV} characteristics of the \ac{VCSEL} measured at room temperature. The \ac{VCSEL} has a maximum output power of $14$~mW, a forward threshold current of about $1.5$~mA, and a roll-over current of $30$~mA. In addition, it provides a linear input dynamic range with a slope efficiency of $0.6$~W/A for forward currents between $5$~mA and $20$~mA, which is suitable for \ac{DCO{-}OFDM} modulation. Fig.~\ref{Fig:VCSEL_OpticalSpectrum} shows the optical spectrum of the \ac{VCSEL} for different values of the \ac{DC} bias current $I_\mathrm{DC}$, measured using a high-resolution \ac{OSA}. It can be observed that when $I_\mathrm{DC}$ increases from $5$~mA to $15$~mA, the peak wavelength shifts slightly from $939.2$~nm to $941.2$~nm. This spectral shift is mainly due to thermal effects, which cause the longitudinal resonance of the cavity to move to longer wavelengths \cite{NLedentsov2022HighSpeed}. For each case, the average wavelength $\bar{\lambda}$ and the \ac{RMS} spectral width $\sigma_\lambda$ are also provided in Fig.~\ref{Fig:VCSEL_OpticalSpectrum}. The \ac{RMS} spectral width is a measure of the effective wavelength spread of the optical spectrum and is defined as the weighted standard deviation of the wavelength around its mean \cite{Sorokin2000SpectralWidth}:
\begin{equation}
    \sigma_\lambda = \sqrt{\dfrac{\int_{0}^{\infty}\big(\lambda-\bar{\lambda}\big)^2 S(\lambda)d\lambda}{\int_{0}^{\infty}S(\lambda)d\lambda}}, \quad \bar{\lambda} = \dfrac{\int_{0}^{\infty}\lambda S(\lambda)}{\int_{0}^{\infty}S(\lambda)d\lambda}.
    \label{Eq:SpectralWidth}
\end{equation}
where $S(\lambda)$ is the emitted power at wavelength $\lambda$. The value of $\sigma_\lambda$ directly determines the \ac{CD} penalty and maximum reachable distance in \ac{MMF} links \cite{NLedentsov2022HighSpeed}. According to Fig.~\ref{Fig:VCSEL_OpticalSpectrum}, by increasing $I_\mathrm{DC}$ from $5$~mA to $15$~mA, $\sigma_\lambda$ increases from $0.22$~nm to $0.591$~nm. As a reference for comparison, the IEEE 802.3 standard for Ethernet over \ac{MMF} links specifies a maximum \ac{RMS} spectral width of $0.65$~nm for \acp{VCSEL} operating at $842$~nm to $948$~nm to reach link distances up to $100$~m \cite{IEEE8023db}. Fig.~\ref{Fig:VCSEL_FarFieldIntensity} shows the \ac{FF} intensity profile of the \ac{VCSEL} measured at different values of $I_\mathrm{DC}$, which confirms that the \ac{VCSEL} has a nearly Gaussian beam profile, with a \ac{FWHM} divergence angle ranging between $15^\circ$ and $18^\circ$ when increasing $I_\mathrm{DC}$ from $5$~mA to $15$~mA. The corresponding divergence half-angle, defined for the beam spot containing $86\%$ of the transmitted power, varies between $12.7^\circ$ and $15.3^\circ$. Furthermore, the \ac{VCSEL} has a $3$-dB modulation bandwidth of $18$~GHz.

\subsection{End-to-End System Setup}
\begin{figure}[!t]
	\centering
	\includegraphics[width=\linewidth]{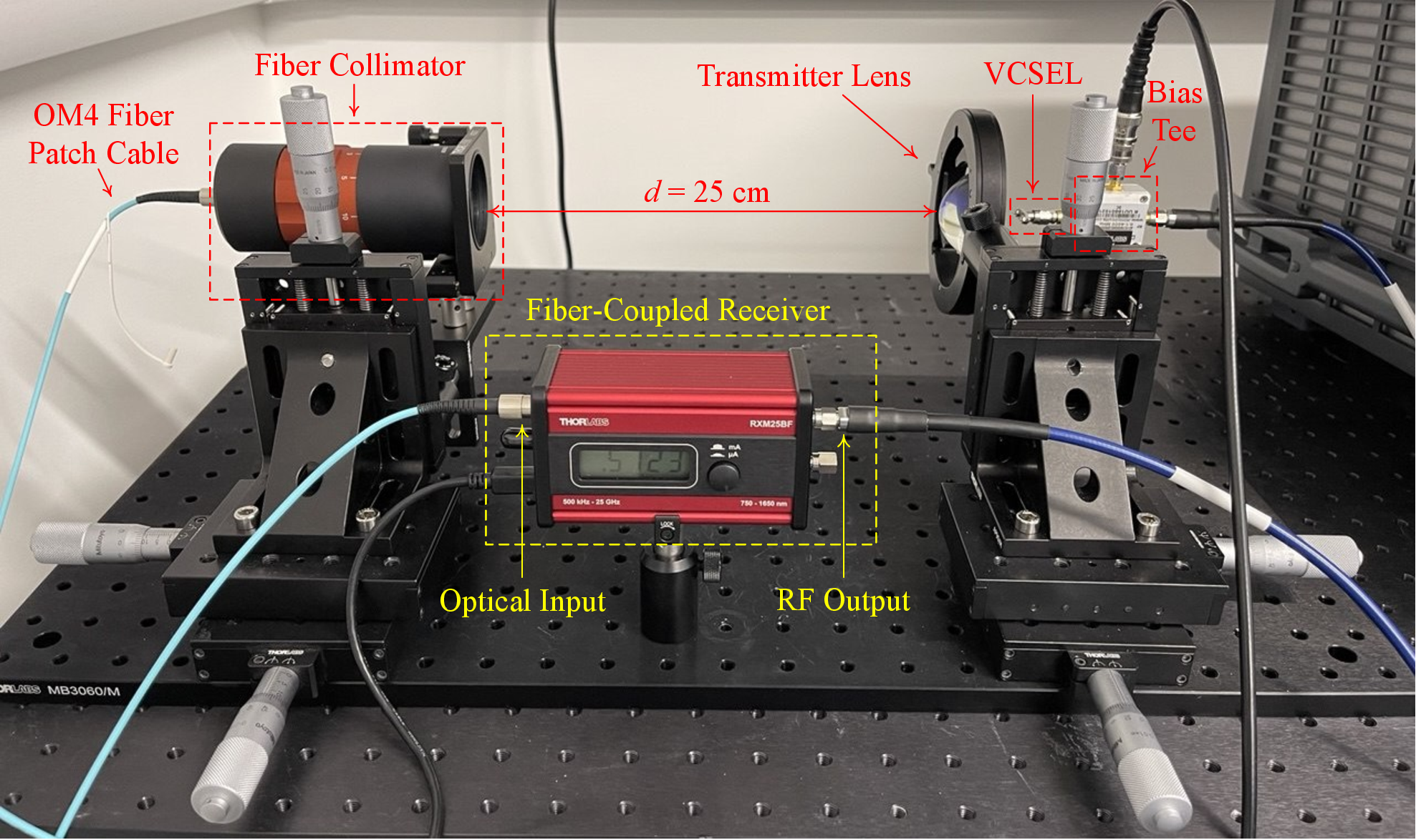}
	\caption{Experimental setup for the VCSEL-based optical wireless link using a fiber-coupled receiver.}
    \label{Fig:ExperimentalSetup}
\end{figure}

Fig.~\ref{Fig:ExperimentalSetup} shows the link configuration including the optical and optomechanical assemblies of the transmitter and receiver. Fig.~\ref{Fig:BlockDiagram} depicts the complete experimental setup. In this work, considering the low-pass and frequency-selective characteristics of the \ac{VCSEL}-based \ac{OWC} system, we use \ac{DCO{-}OFDM} with adaptive subcarrier bit and power allocation to ensure high spectral efficiency. The modulation and demodulation of the \acs{OFDM} signal are implemented based on $N_\mathrm{FFT}$-point \ac{IFFT} and \acs{FFT} operations, respectively, where $N_\mathrm{FFT}=1024$, using offline \ac{DSP} in MATLAB running on a laptop computer. At the transmitter, the information bits are mapped to $M$-ary \ac{QAM} symbols to be loaded across a total of $N_\mathrm{SC}=\frac{N_\mathrm{FFT}}{2}-1=511$ data-carrying subcarriers, with the constellation size reaching up to $M=1024$, depending on the available \ac{SNR}. The \ac{DCO{-}OFDM} frame is constructed by concatenating the block of the complex \ac{QAM} symbols with its conjugated version according to a Hermitian symmetry. A \ac{CP} of length $N_\mathrm{CP}=15$ is then added and an \ac{IFFT} is performed on the resulting frame to generate the time domain \acs{OFDM} signal. This signal is further shaped using a root-raised cosine filter with a roll-off factor of 0.1. The pulse-shaped digital waveform is converted into a wideband analog signal via the Keysight M8195A \ac{AWG}, operating at a sampling frequency of $F_\mathrm{s}=32$~GSa/s. According to the Nyquist-Shannon theorem, the bandwidth $B$ of the signal is related to the sampling frequency via $F_\mathrm{s}\geq2B$. In practice, the signal bandwidth at the \ac{AWG} output can be controlled using the number of \ac{SPS} parameter, denoted by $N_\mathrm{SPS}$, based on $B=\left(F_\mathrm{s}/2\right)/N_\mathrm{SPS}$. We choose $N_\mathrm{SPS}=1$ to obtain $B=16$~GHz. The \ac{AWG} output signal is combined with the required \ac{DC} bias using a Mini Circuits ZX85-12G-S+ \ac{BT} to drive the \ac{VCSEL}. This \ac{BT} has a $12$~GHz bandwidth with a minimum operating frequency of $200$~kHz, and an \ac{IL} of $\mathrm{IL}<1.5$~dB. The wideband \acs{OFDM} signal is fed into the \ac{BT} using a \ac{RF} coaxial cable with a bandwidth of $26$~GHz, a length of $91.44$~cm, and $\mathrm{IL}<1.8$~dB over the desired bandwidth. To collimate the light beam at the transmitter, a Thorlabs ACL5040U-B \acl{ACL} is placed in front of the \ac{VCSEL}. This lens has a diameter of $50$~mm with a \ac{CA} of $45$~mm, a focal length of $40$~mm, a \ac{NA} of $0.6$, and an \ac{ARC} for $650$--$1050$~nm. Note that when the signal is directly modulated on the forward current of the \ac{MA} \ac{VCSEL} for \ac{IM}, although the output beams of the four apertures are overlapping while propagating in free space, there is no crosstalk between them, since they carry the same signal and their signal contributions are constructively added together upon \ac{DD} at the receiver.

At the receiver side, a Thorlabs C80FC-B \ac{AFC} is employed to collect the incident beam from the transmitter located at a $25$~cm distance. This \ac{AFC} has a \ac{CA} of $42.5$~mm, an effective focal length of $80$~mm, an \ac{ARC} for $650$--$1050$~nm matching the coating of the transmitter lens, and a \ac{FC{/}PC} fiber-optic output with a maximum fiber \ac{NA} of $0.25$. As shown in Fig.~\ref{Fig:ExperimentalSetup}, the \ac{AFC} is mounted on a three-axis kinematic mount with high-resolution angular adjusters along with an $\mathrm{XYZ}$ coordinates translation stage, providing six \ac{DOF} required for precise alignment of the focused beam spot on the fiber plane to ensure high coupling efficiency. The output port of the \ac{AFC} is connected to the optical input of the fiber-coupled receiver using a graded-index OM4 \ac{MMF} patch cable of length $1$~m with a core diameter of $50$~{\textmu}m and \ac{FC{/}PC} connectors. We use a Thorlabs RXM25BF fiber-optic photoreceiver, which is equipped with an InGaAs {PIN} \ac{PD} followed by integrated \ac{TIA} and \ac{VGA} stages. This receiver has a broadband responsivity in $750$--$1650$~nm, and a photodetection bandwidth of up to $25$~GHz with a minimum operating frequency of $500$~kHz. The built-in \ac{VGA} helps with adjusting the transimpedance gain and the peak of the receiver frequency response to optimize the performance. There is also a digital screen on the photoreceiver to monitor the average detected photocurrent, which helps with performing the alignment using an unmodulated laser beam. The output analog signal from the photoreceiver is converted to a digital signal by the Keysight UXR0104B oscilloscope at a sampling frequency of $32$~GSa/s. The remaining \ac{DSP} is performed in MATLAB. This includes frame synchronization, \acs{FFT}, single-tap equalization, and \ac{QAM} demapping. For each experiment, a total of $450$ \acs{OFDM} frames consisting of $150$ pilot frames and $300$ data frames are sent through the \ac{OWC} system to measure the \ac{BER} performance. As shown in Fig.~\ref{Fig:ExperimentalSetup}, the back-to-back reference signal provides a copy of the transmitted \acs{OFDM} signal on the oscilloscope Channel~2 through a cable connection with the \ac{AWG}, which can be used as the triggering channel to capture the waveform of the received \acs{OFDM} signal.

Since the \ac{AFC} is mounted by using an optomechanical assembly with six \ac{DOF}, any misalignment including lateral, angular, or axial, has an impact on the light coupling into the \ac{MMF} core. It is particularly sensitive to angular misalignments when adjusting the three tilt angles of the \ac{AFC}. Even a tiny turn of one of the adjusting knobs on the kinematic mount can greatly shift the beam spot on the focal plane of the \ac{AFC}, causing the coupling efficiency to drop dramatically, which makes the alignment process practically challenging. We choose a link distance of $25$~cm to alleviate the complexity of alignment and achieve reliable performance. However, the proposed setup can be systematically reconfigured to increase the link distance. Note that the end-to-end performance is not limited by fiber dispersion or attenuation due to the short length of the OM4 \ac{MMF} cable. To elaborate, Appendix~\ref{Appendix_A} provides the maximum bandwidth and the maximum reachable distance of \ac{MMF} links under both \ac{CD} and \ac{MD}. The fiber patch cable has an \ac{EMB} of $4700$~MHz$\cdot$km and a loss of $\alpha\leq2.3$~dB/km at $850$~nm \cite{ThorlabsOM4MMF}. We adopt these values to present a conservative estimate of the performance at $940$~nm, since \ac{EMB} increases and $\alpha$ decreases with wavelength \cite{Agrawal2010FiberOpticCommunication}. Based on \eqref{Eq:3dB_Bandwidth_Max}, by substituting $L=1$~m, $|D|=65$~ps/nm$\cdot$km \cite{IEEE8023cd_WBMMF}, and $\sigma_\lambda=0.351$~nm from Fig.~\ref{Fig:VCSEL_OpticalSpectrum}, we obtain $f_\mathrm{3dB}\approx3.654$~THz. Compared to the signal bandwidth $B=16$~GHz, $f_\mathrm{3dB}\gg B$, indicating that \ac{CD} and \ac{MD} have practically no impact on the performance. Alternatively, based on \eqref{Eq:ReachableDistance_Max}, we find $L_\mathrm{max} \approx 228.3~\text{m}\gg 1$~m, which leads to the same conclusion. In addition, the fiber attenuation for $L=1$~m is $\alpha L\leq2.3\times10^{-3}$~dB, which is quite negligible.

\subsection{Achievable Data Rate}
In the proposed setup, the system components are carefully selected to support wideband signal transmission. In order to evaluate the system bandwidth, we need to take into account every single component or connection along the \ac{RF} signal path in the end-to-end system. The system bandwidth is limited by the one that has the lowest bandwidth. However, the use of adaptive bit and power loading allows for the signal bandwidth to be extended beyond the $3$-dB bandwidth of the system, due to the allocation of bits to high-frequency subcarriers with low \ac{SNR} values. The data rate of the \ac{DCO{-}OFDM} system is computed by \cite{ESarbazi2022Design2}:
\begin{equation}
    R = \frac{2B}{N_\mathrm{FFT}+N_\mathrm{CP}}\sum_{n=1}^{N_\mathrm{SC}} b_k
    \label{Eq:Rate}
\end{equation}
where $b_k$ is the number of bits allocated to the $k$th subcarrier such that $b_k=\log_2 M_k$ for $M_k>0$ or $b_k=0$ depending on the available \ac{SNR}, with $M_k$ denoting the \ac{QAM} constellation size for the $k$th subcarrier. To meet a given \ac{BER} requirement, the maximum value of $M_k$ is obtained as \cite{Goldsmith1997}:
\begin{equation}
    M_k \leq 1+\frac{\mathrm{SNR}_k}{\Gamma},
    \label{Eq:ConstellationSize}
\end{equation}
where $\mathrm{SNR}_k$ represents the corresponding subcarrier \ac{SNR}, and $\Gamma$ is defined as the \ac{SNR} gap, which is given by \cite{Goldsmith1997}:
\begin{equation}
    \Gamma = \frac{-\ln\left(5\mathrm{BER}\right)}{1.5}.
    \label{Eq:SNR_Gap}
\end{equation}
Hence, the achievable data rate in \eqref{Eq:Rate} is upper-bounded as:
\begin{equation}
    R \leq \frac{2B}{N_\mathrm{FFT}+N_\mathrm{CP}}\sum_{k=1}^{N_\mathrm{SC}} \log_2\left(1+\frac{\mathrm{SNR}_k}{\Gamma}\right).
    \label{Eq:RateUpperBound}
\end{equation}

\begin{table}[t!]
	\centering
	\caption{DC Operating Configurations}
    \begin{threeparttable}
	\begin{tabular}{l|c|c|c|c}
		              & $V_\mathrm{DC}$~[V] & $I_\mathrm{DC}$~[mA] & $P_\mathrm{t}$~[mW] & $P_\mathrm{r}$~[{\textmu}W] \\ \hline
		Config.~I       & $2.40$              & $8.42$               & $4.26$              & $445$   \\
		Config.~II      & $2.44$              & $9.46$               & $4.95$              & $510$   \\ \hline
	\end{tabular}
    \begin{tablenotes}
        \item[] $P_\mathrm{t}$: Transmitted optical power.
    \end{tablenotes}
    \end{threeparttable}
	\label{Tab:DC_Bias_Configs}
\end{table}

\addtocounter{figure}{1}
\begin{figure}[!t]
	\centering
	\includegraphics[width=\linewidth]{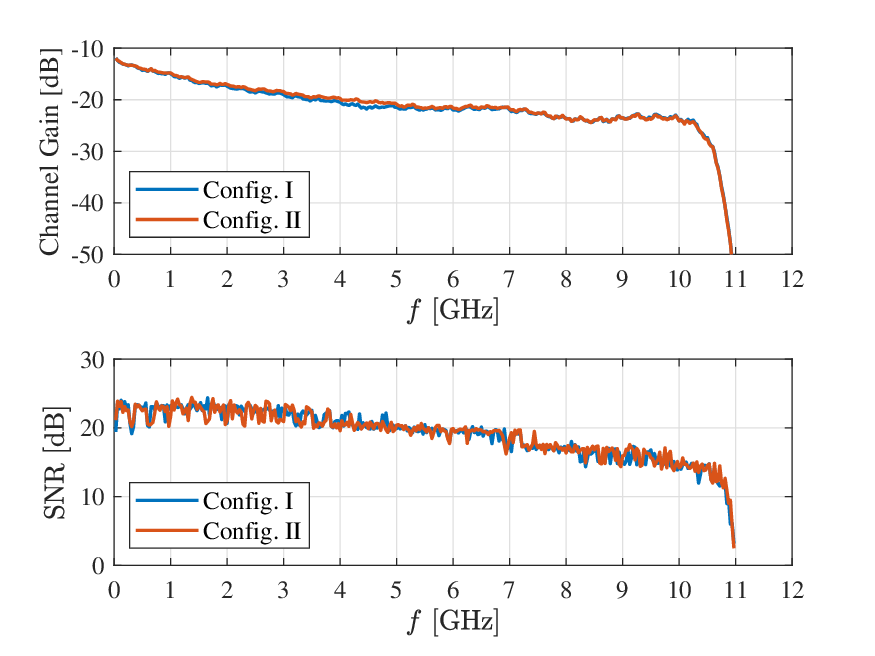}
	\caption{The system frequency response (top figure) and the received electrical SNR (bottom figure) vs frequency.}
	\label{Fig:FrequencyResponse}
\end{figure}

\section{Measurement Results} \label{Sec:Results}
We present the measurement results for the performance of the \ac{VCSEL}-based \ac{OWC} system in terms of the received \ac{SNR}, average \ac{BER} and link data rate. The channel transfer function is estimated by sending pilot $4$-\ac{QAM} symbols of unit variance on all data-carrying subcarriers \cite{CChen2024100Gbps}. The estimated channel information is then used to determine the optimal average transmit power and constellation order of \ac{QAM} symbols for each subcarrier based on the corresponding subcarrier \ac{SNR}, thus adapting the transmission rate and maximizing the system spectral efficiency. In this work, the \ac{HH} bit and power loading algorithm is applied \cite{HughesHartogs1989patent}, which is widely used in multi-carrier communication systems. The optimal \ac{DC} operating point of the \ac{VCSEL} is obtained through extensive measurements and two configurations are identified. These configurations along with the resulting average received optical powers, denoted by $P_\mathrm{r}$, are given in Table~\ref{Tab:DC_Bias_Configs}. In addition, to ensure operation within the linear dynamic range of the \ac{VCSEL}, the peak-to-peak amplitude (i.e., standard deviation) of the \acs{OFDM} signal is set to $V_\mathrm{pp}=0.55$~V. This value is acquired by sweeping $V_\mathrm{pp}$ in an appropriate range according to the \ac{LIV} characteristics of the \ac{VCSEL} and specifying the value that maximizes the link data rate by post-processing the measurement data in MATLAB.

\subsection{SNR Frequency Response: Adaptive Bit and Power Loading}
Fig.~\ref{Fig:FrequencyResponse} shows the channel gain and the electrical \ac{SNR} as a function of frequency. This channel gain represents the channel transfer function which is measured in the frequency domain between the \ac{IFFT} input at the transmitter and the \acs{FFT} output at the receiver. It can be observed that the channel gain exhibits a sharp decline after $10$~GHz. This is primarily caused by the hard cutoff frequency of the oscilloscope at about $11$~GHz. Although the \ac{VCSEL} and the fiber-coupled receiver can reach a higher bandwidth (i.e., $>18$~GHz and $25$~GHz, respectively), in this experiment, the oscilloscope limits the bandwidth of the end-to-end system. We note that since the \ac{BT} has a bandwidth of $12$~GHz, it also causes a roll-off on the frequency response at high frequencies. However, the cutoff frequency at $11$~GHz is still dominant. Evidently, the resulting \ac{SNR} values are above $0$~dB over the entire range of frequencies up to $11$~GHz. Thus, the effective modulation bandwidth that can be used for data transmission on \acs{OFDM} subcarriers is $11$~GHz. It can be observed that $20~\text{dB}\leq\mathrm{SNR}<25~\text{dB}$ for $f\leq6$~GHz, $10~\text{dB}\leq\mathrm{SNR}<20$~dB for $6~\text{GHz}<f\leq10.8$~GHz, and $0~\text{dB}<\mathrm{SNR}<10$~dB for $10.8~\text{GHz}<f\leq11$~GHz. Furthermore, there are oscillations in the \ac{SNR} response especially at frequencies lower than $4$~GHz, due to the impedance mismatch between the \ac{VCSEL} and the \ac{BT}. In fact, while the \ac{BT} has an output impedance of $50$~{\textOmega}, the input impedance of the \ac{VCSEL} is significantly less than $50$~{\textOmega} (i.e., about $25$~\textOmega). Although the \ac{VCSEL} chip is packaged on an standard \ac{RF}-V connector and it is directly connected to the \ac{BT} output, as shown in Fig.~\ref{Fig:ExperimentalSetup}, its low input impedance leads to signal reflections at this critical connection, resulting in \ac{RF} interference that degrades the \ac{SNR} and hence oscillations appear in the \ac{SNR} frequency response. The remaining cable connections along the \ac{RF} signal path at the transmitter and receiver are carefully chosen to be as short as possible to minimize \ac{RF} signal distortions.

\begin{figure}[!t]
	\centering
	\includegraphics[width=\linewidth]{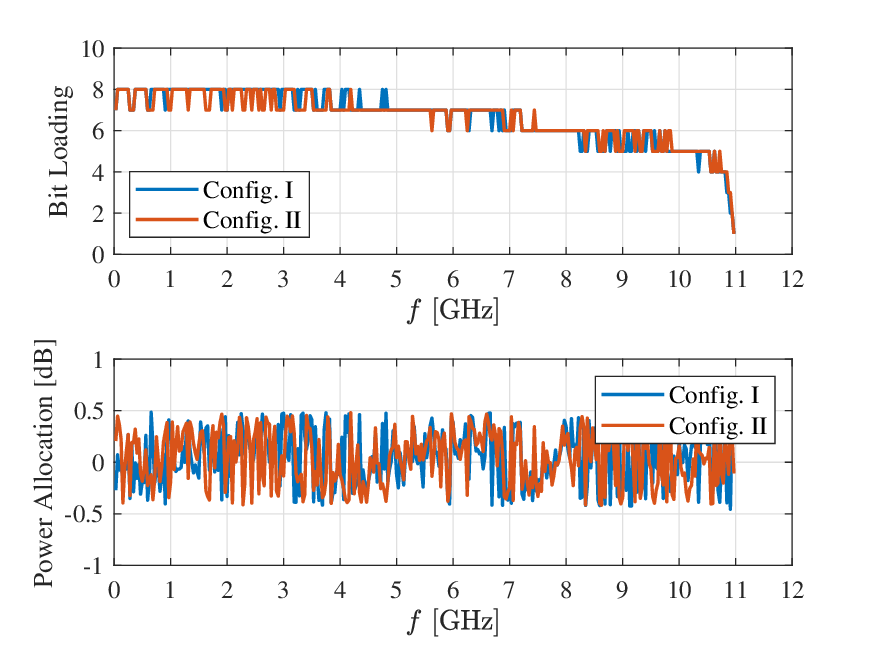}
	\caption{Adaptive bit loading (top figure) and power allocation (bottom figure) results vs subcarrier frequency for $R>70$~Gb/s, under $\mathrm{BER}=3.3\times10^{-2}$ for Config.~I, and $\mathrm{BER}=3.1\times10^{-2}$ for Config.~II.}
	\label{Fig:BitLoading}
    \vspace{-5pt}
\end{figure}

The estimated channel gain and \ac{SNR}, as shown in Fig.~\ref{Fig:FrequencyResponse}, are used to allocate the optimal number of bits and amount of power to each subcarrier. Fig.~\ref{Fig:BitLoading} shows an example of adaptive bit loading and power allocation results. It can be observed that the number of loaded bits is greatly dependent on the estimated \ac{SNR} across the operating bandwidth. Also, adaptive power allocation balances out the \ac{SNR} level to ensure that it is adequate to send the allocated number of bits at the required \ac{BER}. It can be observed that $7$ to $8$ bits are loaded on low frequency subcarriers up to $6$~GHz, $4$ to $6$ bits are loaded to high frequency subcarriers between $6$~GHz and $10.8$~GHz, and $1$ to $3$ bits are carried by the remaining subcarriers between $10.8$~GHz and $11$~GHz.

\subsection{BER vs Achievable Data Rate}
\begin{figure}[!t]
	\centering
	\includegraphics[width=0.9\linewidth]{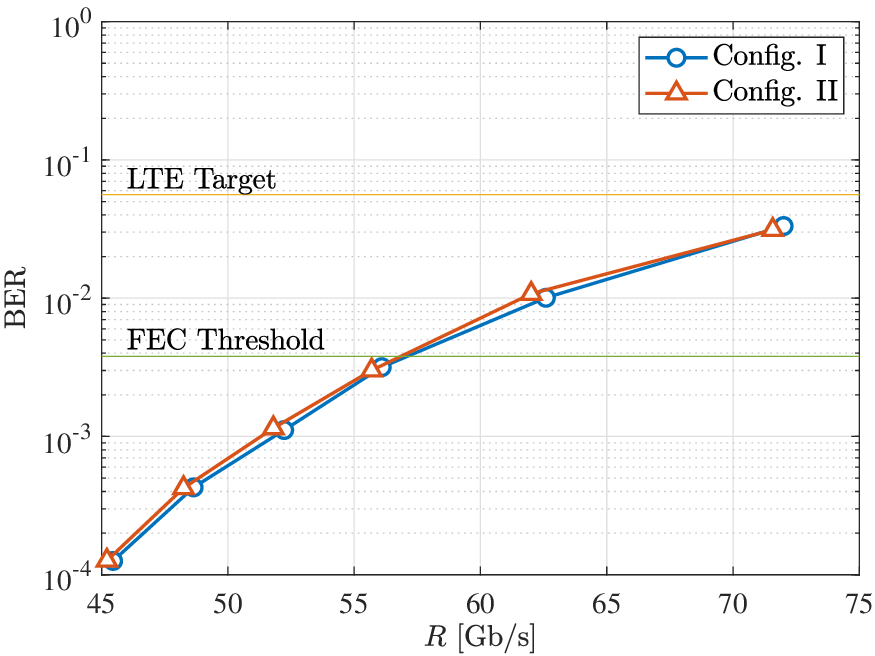}
	\caption{The BER vs achievable data rate performance for the VCSEL-based optical wireless link.}
	\label{Fig:BER_Rate}
\end{figure}

Fig.~\ref{Fig:BER_Rate} shows the average \ac{BER} performance with respect to the achievable data rate. It can be observed that the achievable data rate increases steadily at the cost of increasing the \ac{BER} at the same time. It can also be observed that Config.~I performs slightly better than Config.~II. A possible explanation for this is Config.~I sets the \ac{VCSEL}'s \ac{DC} bias at a lower operating point with slightly smaller voltage and current values. Therefore, the modulating signal experiences less nonlinear distortion effects, particularly for the signal swing above the \ac{DC} operating point. In other words, in Config.~I, the \acs{OFDM} signal variations match the linear dynamic range of the \ac{VCSEL}. As a result, the \ac{SNR} is slightly improved especially in the lower frequency region up to $6$~GHz, as shown in Fig.~\ref{Fig:FrequencyResponse}, thereby improving the bit loading performance, as shown in Fig.~\ref{Fig:BitLoading}. We consider a target \ac{BER} of $5.6\times10^{-2}$ according to system-level requirements in \ac{LTE} systems. In fact, in \ac{LTE} systems, link adaptation aims to maintain a \ac{BLER} of $10\%$ after decoding \cite[Chapter~10, pp.~215--248]{Stefania2011LTE}. This allows the system to tolerate a relatively high raw, uncoded \ac{BER} on the order of $10^{-2}$ to $10^{-1}$ at the physical layer depending on the selected \ac{MCS} and the operating \ac{SNR} so as to maximize spectral efficiency, relying on advanced \ac{FEC} coding schemes such as Turbo codes or \ac{HARQ} mechanisms to enhance the \ac{BER} performance \cite{3GPP36.213}. The final post-\ac{FEC} \ac{BER} is expected to be at $10^{-6}$ or lower after decoding with $3\%$ to $5\%$ overhead \cite{Smith2012Staircase}. The \ac{BER} performance can be further reduced to $10^{-15}$ after decoding by using staircase codes with a $6.25\%$ overhead \cite{Smith2012Staircase}. From Fig.~\ref{Fig:BER_Rate}, it can be observed that the achievable data rate exceeds $70$~Gb/s while $\mathrm{BER}<5.6\times10^{-2}$. More precisely, the \ac{VCSEL}-based optical wireless link achieves data rates of $R=72$~Gb/s and $R=71.6$~Gb/s at $\mathrm{BER}=3.3\times10^{-2}$ and $\mathrm{BER}=3.1\times10^{-2}$ under Config.~I and Config.~II, respectively. We note that both configurations achieve a data rate above $57$~Gb/s at a pre-\ac{FEC} \ac{BER} level of $3.8\times10^{-3}$, which offers a $50\%$ improvement relative to the data rate reported in \cite{Osahon2024VCSEL38Gbps} at the  same \ac{BER} level.

\subsection{Achieving $>\mathit{100}$~Gb/s Through SNR Extrapolation}
\begin{figure}[!t]
    \centering
    \subfloat[Config. I \label{Fig:SNR_Extrapolation_Config_I}]{\includegraphics[width=0.9\linewidth, keepaspectratio=true]{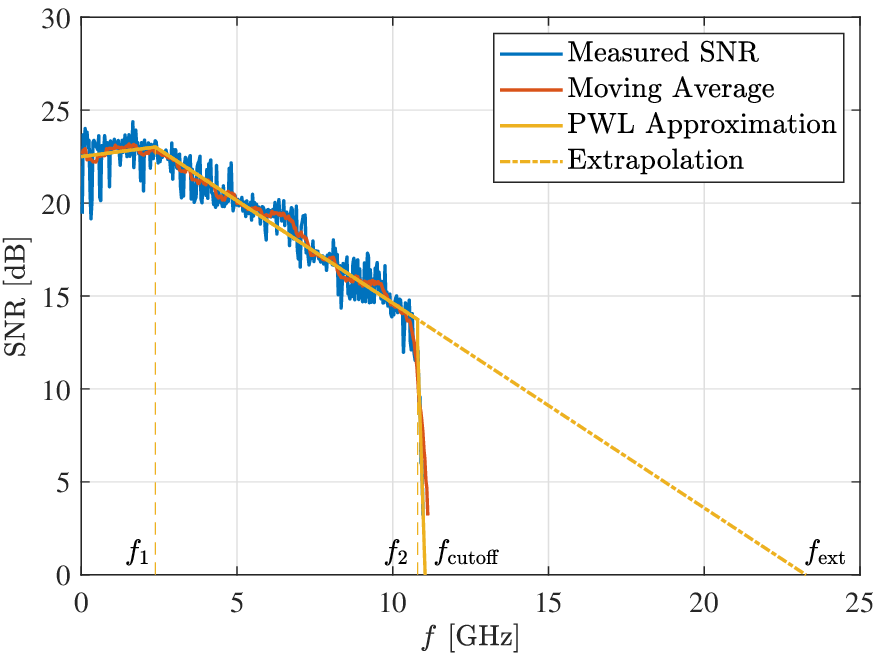}} \vspace{-1pt}
    \subfloat[Config. II \label{Fig:SNR_Extrapolation_Config_II}]{\includegraphics[width=0.9\linewidth, keepaspectratio=true]{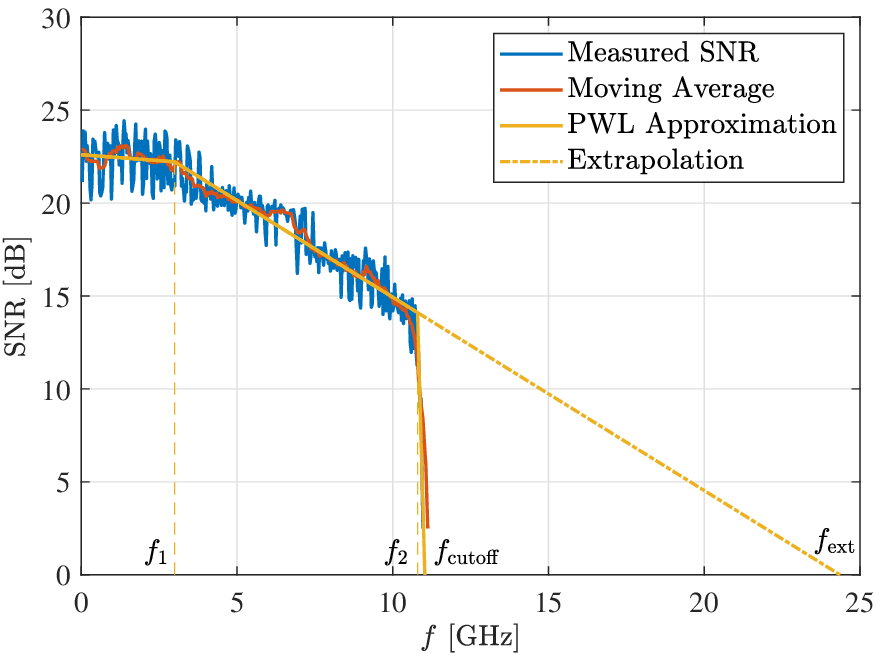}}
    \caption{Extrapolating the SNR frequency response beyond the cutoff frequency of the oscilloscope.}
    \label{Fig:SNR_Extrapolation}
\end{figure}

Looking at Fig.~\ref{Fig:FrequencyResponse}, a natural question arises as to how much performance enhancement can be achieved if the oscilloscope bandwidth constraint is eliminated? To address this question, we estimate the trend of the \ac{SNR} in the frequency domain by approximating the measured \ac{SNR} frequency response based on \ac{PWL} regression. For each configuration, the approximated \ac{SNR} is characterized by two frequencies of $f_1$ and $f_2$ such that $f_1<f_2<f_\mathrm{cutoff}$, where $f_\mathrm{cutoff}=11$~GHz indicates the frequency at which the \ac{SNR} drops to $0$~dB, as shown in Fig.~\ref{Fig:SNR_Extrapolation}. A moving average filter with a window size of $10$ is also employed to smooth out the ripples in the measured \ac{SNR} results due to \ac{RF} reflections. The resulting curve manifests the \ac{SNR} trend and is used as a guide for line fitting in each frequency band. The computed values of $f_1$ and $f_2$ for both configurations are listed in Table~\ref{Tab:PWL_Approximation_Frequencies}. The approximated \ac{SNR} is then extrapolated by removing the hard limit of the oscilloscope at $f=f_\mathrm{cutoff}$ and letting the \ac{SNR} to follow the same decreasing trend as in the frequency band $f_1\leq f<f_2$ until it crosses $0$~dB at $f=f_\mathrm{ext}$, as shown in Fig.~\ref{Fig:SNR_Extrapolation}. This leads to $f_\mathrm{ext}=23.26$~GHz for Config.~I, and  $f_\mathrm{ext}=24.36$~GHz for Config.~II. The rationale behind this extrapolation lies in the fact that the frequency response of \ac{SM} \ac{MA} \acp{VCSEL} exhibits a monotonically decreasing behavior with a consistent roll-off at high frequencies \cite{Chorchos2025MAVCSELs}. Thus, when there is no other factor limiting the end-to-end modulation bandwidth, the system frequency response follows a similar decreasing trend.

\begin{table}[!t]
	\centering
	\caption{SNR Approximation and Extrapolation Frequencies}
	\begin{tabular}{l|c|c|c}
		              & $f_1$~[GHz]  & $f_2$~[GHz]  & $f_\mathrm{ext}$~[GHz] \\ \hline
		Config.~I       & $2.38$       & $10.80$       & $23.26$                \\
		Config.~II      & $3.00$       & $10.80$       & $24.36$                \\ \hline
	\end{tabular}
	\label{Tab:PWL_Approximation_Frequencies}
\end{table}

\begin{figure}[!t]
	\centering
	\includegraphics[width=0.9\linewidth]{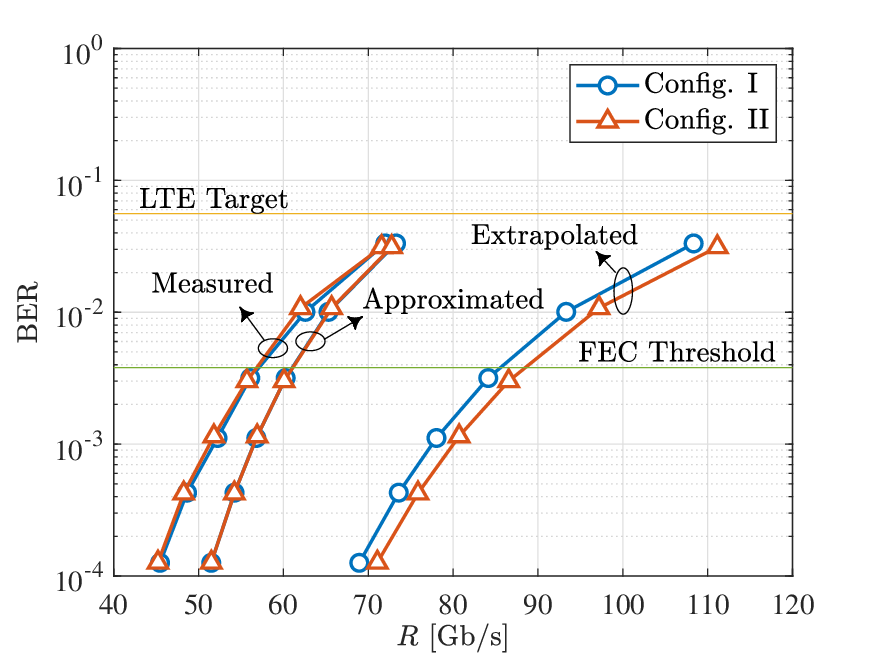}
	\caption{The BER vs achievable data rate performance based on the upper bound in \eqref{Eq:RateUpperBoundIntegral_Extrapolated} using the approximated and extrapolated SNR results in Fig.~\ref{Fig:SNR_Extrapolation}.}
    \label{Fig:BER_Rate_Extrapolated}
\end{figure}

For an \ac{AWGN} channel with an \ac{SNR} frequency response of $\mathrm{SNR}(f)$, the achievable data rate can be upper-bounded as \cite{Forney1998ModulationCoding}:
\begin{equation}
    R \leq \int_{0}^{\infty}\log_2\left(1+\frac{\mathrm{SNR}(f)}{\Gamma}\right)df,
    \label{Eq:RateUpperBoundIntegral}
\end{equation}
where $\Gamma$ is given by \eqref{Eq:SNR_Gap}. When evaluating the upper bound in \eqref{Eq:RateUpperBoundIntegral} for the approximated and extrapolated \ac{SNR}, we have:
\begin{equation}
    R \leq \int_{0}^{f_\mathrm{max}}\log_2\left(1+\frac{\mathrm{SNR}(f)}{\Gamma}\right)df,
    \label{Eq:RateUpperBoundIntegral_Extrapolated}
\end{equation}
where $f_\mathrm{max}$ is replaced by $f_\mathrm{cutoff}$ for the approximated \ac{SNR}, and by $f_\mathrm{ext}$ for the extrapolated \ac{SNR}. Fig.~\ref{Fig:BER_Rate_Extrapolated} shows the \ac{BER} performance with respect to the achievable data rate for both cases based on the upper bound in \eqref{Eq:RateUpperBoundIntegral_Extrapolated}. As observed from Fig.~\ref{Fig:BER_Rate_Extrapolated}, under the approximated \ac{SNR}, the upper bound of the data rate gets tighter for higher \ac{BER} values, indicating that it can be safely used to estimate the maximum achievable data rate. With the extrapolated \ac{SNR}, the achievable data rate is remarkably enhanced, reaching a maximum value of $108.3$~Gb/s and $111.1$~Gb/s at a \ac{BER} of $3.3\times10^{-2}$ and $3.1\times10^{-2}$ for Config.~I and Config.~II, respectively. This confirms that the system can achieve data rates $>100$~Gb/s.

\section{Conclusions} \label{Sec:Conclusions}
In this paper, we have demonstrated an experimental proof-of-concept for ultra-high data rate laser-based optical wireless links for future bandwidth-intensive applications. To this end, we employ a \ac{SM} \ac{MA} \ac{VCSEL} operating at $940$~nm with an ultra-wide modulation bandwidth of $18$~GHz. We also employ a $25$~GHz \ac{MM} fiber-optic photoreceiver to allow the \ac{VCSEL}'s modulation bandwidth to be utilized by applying \ac{DCO{-}OFDM} with adaptive subcarrier bit and power loading. The experimental results corroborate that the proposed system setup achieves a record-high data rate of $72$~Gb/s at a pre-\ac{FEC} \ac{BER} of $3.3\times10^{-2}$ which is below the \ac{LTE} target (i.e., $5.6\times10^{-2}$), while it attains $57$~Gb/s at a pre-\ac{FEC} \ac{BER} of $3.8\times10^{-3}$, over a link distance of $25$~cm. The link range and the received \ac{SNR} can be increased using meticulous alignment procedures to maximize optical coupling efficiency at the \ac{MMF} port of the receiver collimator. Furthermore, the $11$~GHz cutoff frequency of the oscilloscope is currently limiting the end-to-end performance of the system. Otherwise, considering the \ac{VCSEL} bandwidth and the \ac{SNR} response, this system can reach data rates beyond $100$~Gb/s. Through extrapolating the \ac{SNR} response in the absence of the hard cutoff frequency at $11$~GHz, it was verified that this system can potentially achieve $111.1$~Gb/s at a pre-\ac{FEC} \ac{BER} of $3.1\times10^{-2}$. Therefore, more work is yet to be done to realize the full potential of the \ac{SM} \ac{MA} \ac{VCSEL} for ultra-high speed \ac{OWC} system design.

\appendices
\section{Performance Bounds of MMF Under Dispersion} \label{Appendix_A}
The maximum bandwidth and reachable distance of \ac{MMF} links under dispersion are derived, assuming Gaussian impulse responses for both \ac{CD} and \ac{MD} \cite{Agrawal2010FiberOpticCommunication}. The \ac{RMS} delay spread due to \ac{CD} is given by \cite{Agrawal2010FiberOpticCommunication}:
\begin{equation}
    \sigma_\mathrm{CD} = |D|\hspace{1pt}\sigma_\lambda\hspace{1pt}L,
    \label{Eq:RMS_DelaySpread_CD}
\end{equation}
where $D$ is the \ac{CD} coefficient in ps/nm$\cdot$km, $\sigma_\lambda$ is the \ac{RMS} spectral width of the laser source in nm, and $L$ is the fiber length in km. The $3$~dB cutoff frequency for the \ac{CD}-induced pulse broadening is obtained as \cite{Agrawal2010FiberOpticCommunication}:
\begin{equation}
    f_\mathrm{3dB,CD} = \frac{\sqrt{\ln2}}{2\pi \sigma_\mathrm{CD}} = \frac{\sqrt{\ln2}}{2\pi|D|\sigma_\lambda L}.
    \label{Eq:3dB_Bandwidth_CD}
\end{equation}
The $3$~dB cutoff frequency corresponding to the \ac{MD}-induced pulse broadening is \cite{IEEE8023db2022}:
\begin{equation}
    f_\mathrm{3dB,MD} = \frac{\mathrm{EMB}}{L},
    \label{Eq:3dB_Bandwidth_MD}
\end{equation}
where the \ac{EMB} of \ac{MMF} cables is typically given in MHz$\cdot$km. The \ac{RMS} delay spread associated with \ac{MD} is then given by:
\begin{equation}
    \sigma_\mathrm{MD} = \frac{\sqrt{\ln2}}{2\pi f_\mathrm{3dB,MD}}.
    \label{Eq:RMS_DelaySpread_MD}
\end{equation}
The total \ac{RMS} delay spread under the joint effect of \ac{CD} and \ac{MD} can be expressed as \cite{Agrawal2010FiberOpticCommunication}: 
\begin{equation}
    \sigma_\mathrm{t} = \sqrt{\sigma_\mathrm{CD}^2+\sigma_\mathrm{MD}^2},
    \label{Eq:RMS_DelaySpread_Total}
\end{equation}
which leads to the combined $3$~dB cutoff frequency of:
\begin{equation}
    f_\mathrm{3dB} = \frac{\sqrt{\ln2}}{2\pi\sigma_\mathrm{t}} = \frac{1}{\sqrt{\dfrac{1}{f_\mathrm{3dB,CD}^2}+\dfrac{1}{f_\mathrm{3dB,MD}^2}}}.
    \label{Eq:3dB_Bandwidth_Total}
\end{equation}
By substituting \eqref{Eq:3dB_Bandwidth_CD} and \eqref{Eq:3dB_Bandwidth_MD} into \eqref{Eq:3dB_Bandwidth_Total}, the maximum bandwidth supported by \ac{MMF} for a running distance of $L$ is derived as:
\begin{equation}
    f_\mathrm{3dB} = \frac{1}{L\sqrt{\dfrac{\left(2\pi|D|\sigma_\lambda\right)^2}{\ln2}+\dfrac{1}{\mathrm{EMB}^2}}}.
    \label{Eq:3dB_Bandwidth_Max}
\end{equation}
The maximum reachable distance $L_\mathrm{max}$ is determined by the upper bound of $L$ such that $f_\mathrm{3dB}\geq B$ based on \eqref{Eq:3dB_Bandwidth_Max}, with $B$ representing the signal bandwidth, resulting in:
\begin{equation}
    L \leq \frac{1}{B\sqrt{\dfrac{\left(2\pi|D|\sigma_\lambda\right)^2}{\ln2}+\dfrac{1}{\mathrm{EMB}^2}}}.
    \label{Eq:ReachableDistance_Max}
\end{equation}
%


\section*{Acknowledgement}
This work was financially supported by the Engineering and Physical Sciences Research Council (EPSRC) under grant EP/Y037243/1 `Platform Driving The Ultimate Connectivity (TITAN)'.

\bibliographystyle{IEEEtran}
\bibliography{IEEEabrv,references}

\begin{IEEEbiography}[{\includegraphics[width=1in,height=1.25in,clip,keepaspectratio]{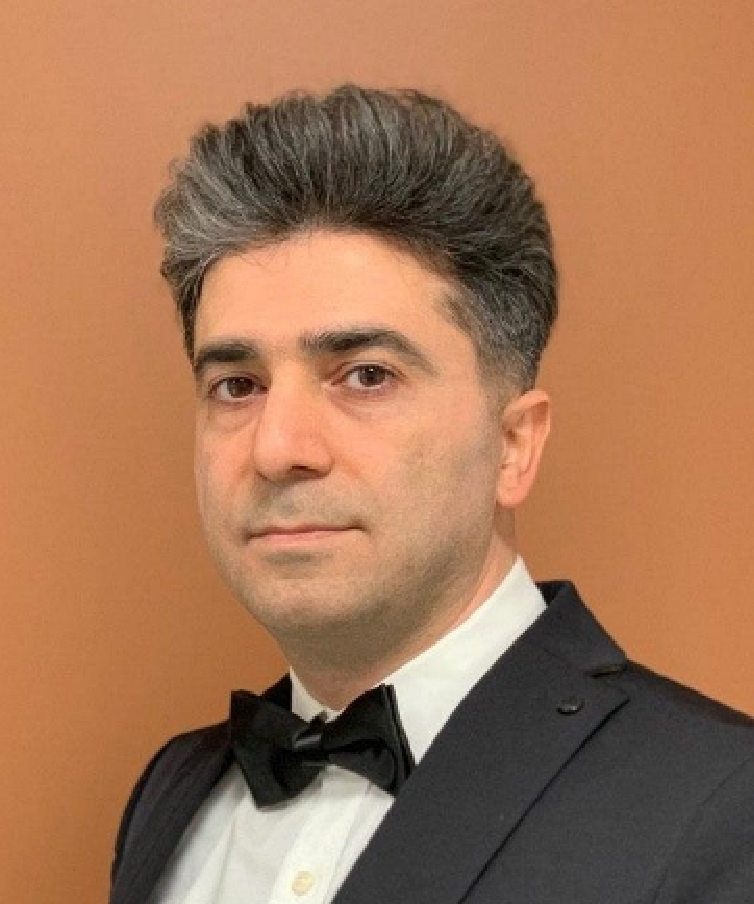}}]{Hossein Kazemi}
(Member, IEEE) received the Ph.D. degree in Electrical Engineering from The University of Edinburgh, U.K., in 2019. He also received the M.Sc. degree in Electrical Engineering (Microelectronic Circuits) from Sharif University of Technology, Tehran, Iran, in 2011, and the M.Sc. degree (Hons.) in Electrical Engineering (Wireless Communications) from Ozyegin University, Istanbul, Turkey, in 2014. He is a Postdoctoral Research Associate at the LiFi Research and Development Center, University of Cambridge, U.K. Dr Kazemi was the recipient of the Best Paper Award for the 2022 IEEE Global Communications Conference (GLOBECOM). His current research interests include the design, analysis and optimization of ultra-high-speed optical wireless communication systems for 6G and beyond networks.
\end{IEEEbiography}

\vfill

\begin{IEEEbiography}[{\includegraphics[width=1in,height=1.25in,clip,keepaspectratio]{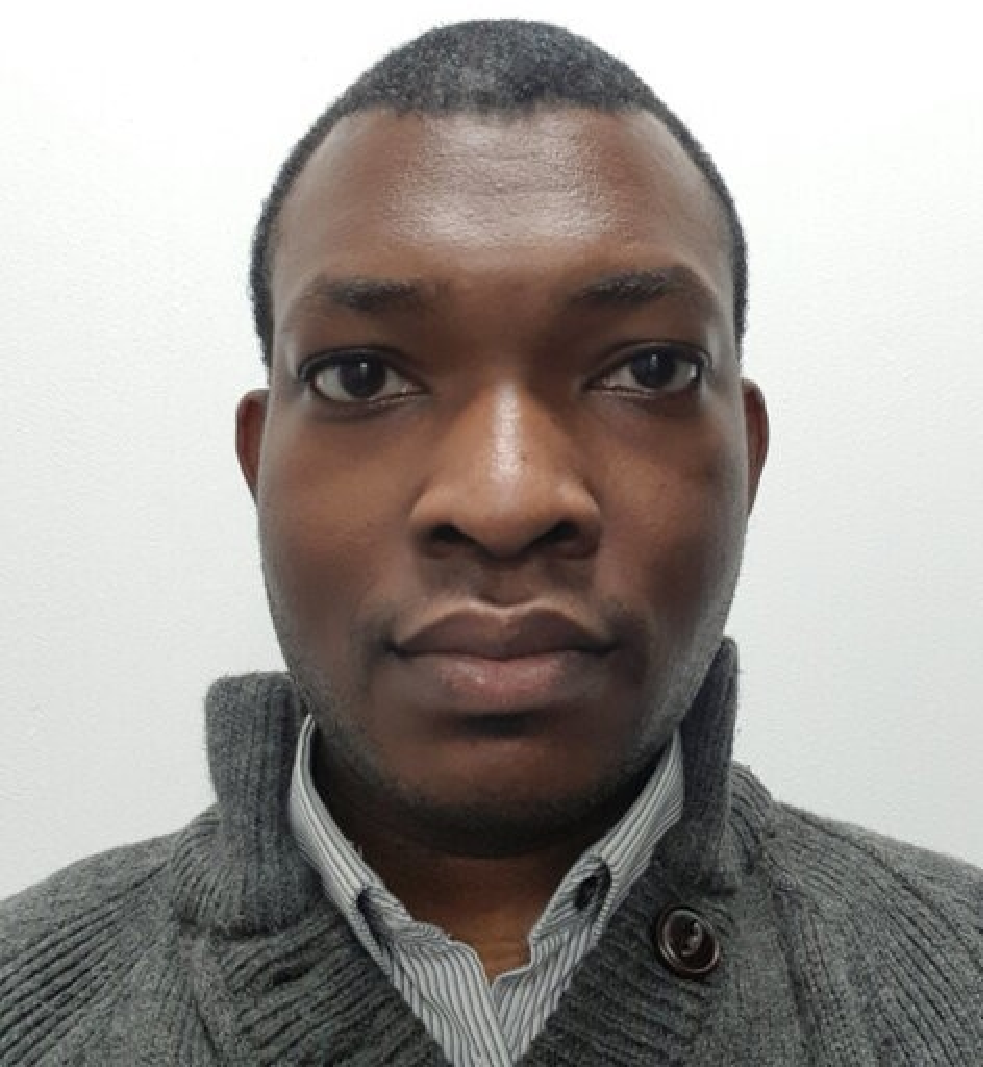}}]{Isaac N. O. Osahon}
(Member, IEEE) received the B.Eng. degree in Electrical and Electronic Engineering (first class hons.) from Covenant University, Ota, Nigeria, in 2012, the M.Sc. degree in Internet Engineering from the University College London, London, in 2015, and the Ph.D. degree at the Institute for Digital Communications, The University of Edinburgh, in 2020. He has worked as a researcher on digital signal processing for optical fiber and wireless communication systems in notable U.K. universities. He is a Postdoctoral Research Associate at the LiFi Research and Development Center, University of Cambridge, U.K. His current research interests include optical communications, advanced modulation schemes, artificial neural networks, digital equalization techniques, visible light positioning and physical layer security.
\end{IEEEbiography}

\vfill

\begin{IEEEbiography}[{\includegraphics[width=1in,height=1.25in,clip,keepaspectratio]{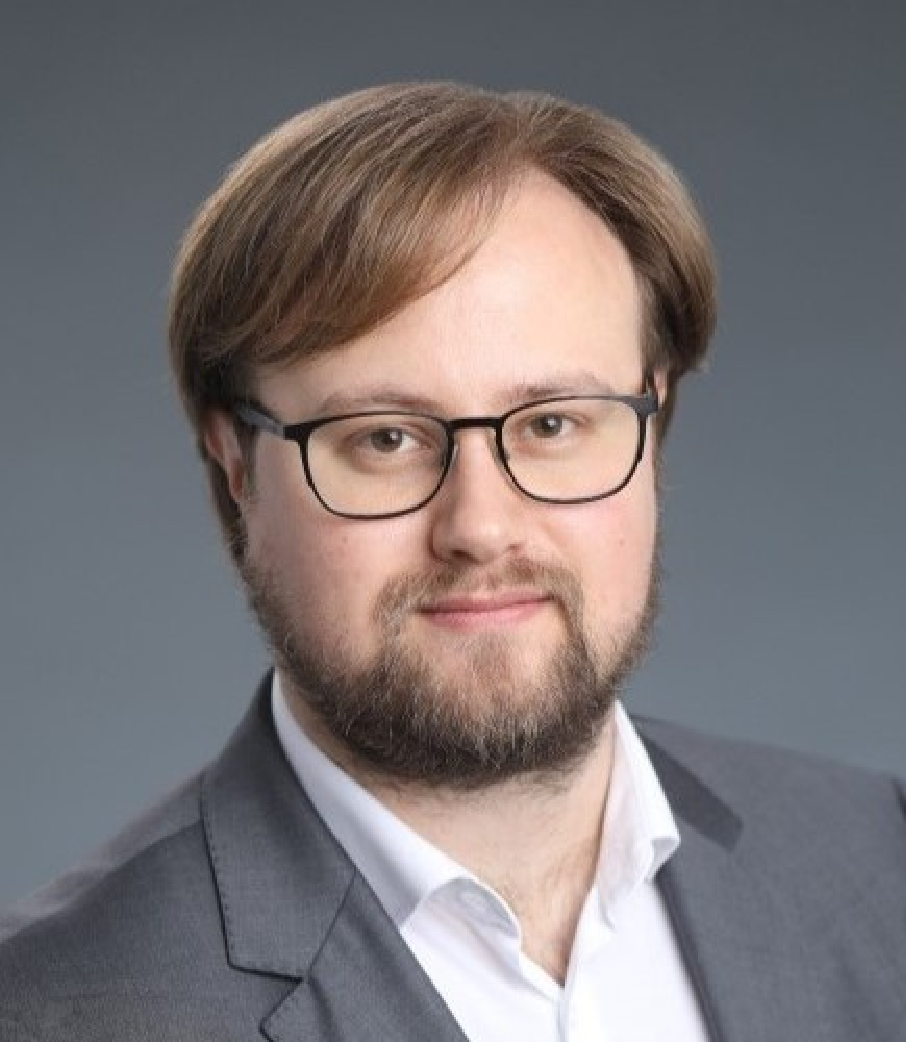}}]{Nikolay Ledentsov Jr.} 
received his B.Sc. and M.Sc. degrees from the Technical University of Berlin, Germany, in 2012 and 2014, respectively, where his research focused on the growth and characterization of Indium Aluminum Gallium Nitride (InAlGaN) green and ultraviolet (UV) light emitting diodes (LEDs). He completed his Ph.D. at the Technical University of Warsaw, Poland, in 2023, focusing on high-speed data transmission with intrared (IR) vertical-cavity surface-emitting lasers (VCSELs). From 2013 to 2024, he was a Senior Engineer at VI Systems GmbH, where he was responsible for the research and development of VCSELs for high-speed optical links, and manufacturing and characterization of light emitters and photodiodes. He is currently Head of Characterization and VCSEL Development at EPIGAP OSA Photonics GmbH.
\end{IEEEbiography}

\vfill

\begin{IEEEbiography}[{\includegraphics[width=1in,height=1.25in,clip,keepaspectratio]{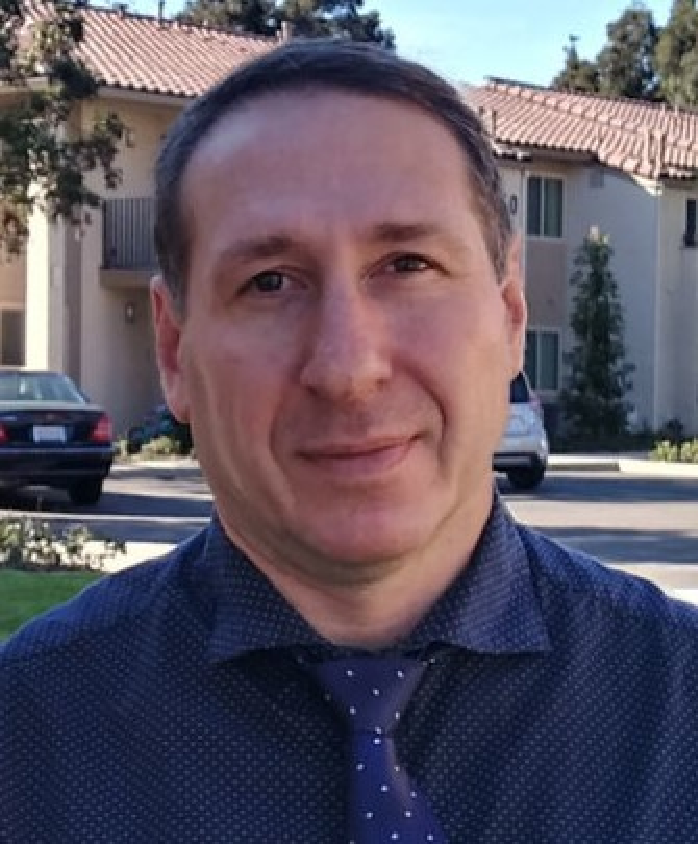}}]{Ilya Titkov} 
received his Ph.D. degree in Semiconductor Manufacturing Technology from Peter the Great St. Petersburg Polytechnic University, Saint Petersburg, Russia, in 2000. From 2002 to 2009, he worked as a researcher at the Ioffe Institute’s Laboratory of Semiconductor Devices Physics in Saint Petersburg, Russia, where he focused on semiconductor material physics and oxide-based device structures. Since 2014, he has held a research position at the Aston Institute of Photonic Technologies, Aston University, Birmingham, U.K., contributing to LED efficiency analysis and photonic device development. From 2019 to 2024, he served as an R\&D Manager and Senior Researcher at VI Systems GmbH, specializing in advanced characterization methods and failure analysis for high-speed VCSEL technologies and optoelectronic devices. He is currently an R\&D Manager at EPIGAP OSA Photonics GmbH.
\end{IEEEbiography}

\vfill

\begin{IEEEbiography}[{\includegraphics[width=1in,height=1.25in,clip,keepaspectratio]{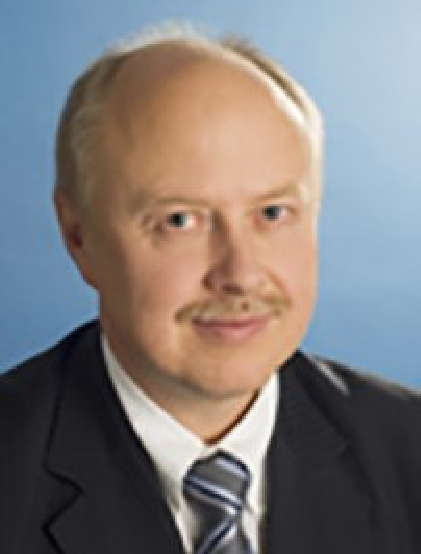}}]{Nikolay Ledentsov} 
(Senior Member, IEEE) received the Diploma in electrical engineer from the Electrical Engineering Institute, Leningrad, Russia, in 1982, the Ph.D. ( Cand. Sci.) and Dr. Sci. (Habil.) degrees in physics and mathematics from A. F. Ioffe Institute, Leningrad/Saint Petersburg, Russia, in 1987 and 1994, respectively. Since 1994, he has been a Professor of electrical engineering and, since 2005, a certified Professor of physics and mathematics with Ioffe Institute. During 1996--2007, he was a Professor with the Technical University of Berlin, Berlin, Germany. He has authored or coauthored more than 900 papers in technical journals and conference proceedings and 38 patent families. His Hirsch factor is 83. His current research interests include physics and technology of semiconductor nanostructures and design and technology of advanced optoelectronic devices. He is a Fellow of the Institute of Physics and a Member of the Russian Academy of Sciences. He was the recipient of the Young Scientist Award from the International Symposium on Compound Semiconductors in 1996 for outstanding contributions to the development of physics and MBE growth of InGaAs/GaAs quantum dots structures and quantum dot lasers, State Prize of Russia for Science and Technology in 2001, Prize of the Berlin Brandenburg Academy of Sciences in 2002, and other awards and recognitions. During 1995--1996, he was the recipient of Alexander von Humboldt Fellowship. Since 2006, he has been the Chief Executive Officer of VI Systems GmbH.
\end{IEEEbiography}

\vfill

\begin{IEEEbiography}[{\includegraphics[width=1in,height=1.25in,clip,keepaspectratio]{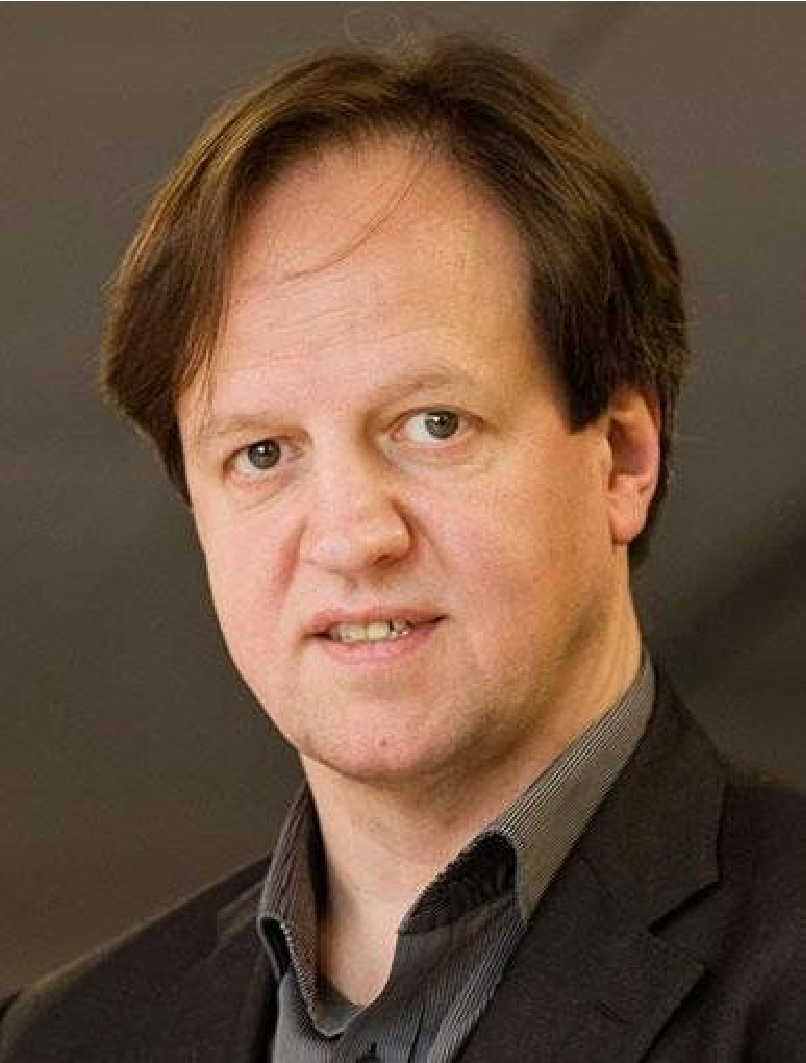}}]{Harald Haas} 
(Fellow, IEEE) received his Ph.D. from The University of Edinburgh, U.K., in 2001. He is the Van Eck Chair of Engineering at the University of Cambridge and the founder of pureLiFi Ltd., where he also serves as the Chief Scientific Officer (CSO). He is the Director of the LiFi Research and Development Centre at the University of Cambridge. His recent research interests focus on photonics, communication theory and signal processing for optical wireless communication systems. Since 2017, he has been recognised as a highly cited researcher by Clarivate/Web of Science. He has delivered two TED talks and one TEDx talk. In 2016, he received the Outstanding Achievement Award from the International Solid State Lighting Alliance. He was awarded the Royal Society Wolfson Research Merit Award in 2017, the IEEE Vehicular Technology Society James Evans Avant Garde Award in 2019, and the Enginuity: The Connect Places Innovation Award in 2021. In 2022, he received the Humboldt Research Award for his research contributions. He is a Fellow of the Royal Academy of Engineering (RAEng), the Royal Society of Edinburgh (RSE), and the Institution of Engineering and Technology (IET). In 2023, he was shortlisted for the European Inventor Award.
\end{IEEEbiography}

\end{document}